\documentclass{aa} 
\usepackage{graphicx}
\DeclareGraphicsExtensions{.eps,.ps}
\usepackage{txfonts}
\usepackage{subfigure}
\usepackage{float}
\usepackage{natbib}
\bibpunct{(}{)}{;}{a}{}{,}
\begin{document}
   \title{An XMM-\textsl{Newton} study of Hyper-Luminous Infrared Galaxies}


   \author{A. Ruiz
          \and
	  F.J. Carrera
		  \and
	  F. Panessa
          \inst{}
         }

   \institute{Instituto de F\'isica de Cantabria (IFCA), CSIC-UC,
              Santander 39005,  Spain\\
              \email{ruizca@ifca.unican.es}
             }

   \date{Received November 7, 2006; accepted May 18, 2007}


  \abstract 
   {} 
    {Hyper-Luminous Infrared Galaxies (HLIRGs) are the most luminous
    persistent objects in the Universe. They exhibit extremely high
    star formation rates, and most of them seem to harbour an Active
    Galactic Nucleus (AGN). They are unique laboratories to
    investigate utmost star formation, and its connection to
    super-massive black hole growth. X-ray studies of HLIRGs have the
    potential to unravel the AGN contribution to the bolometric output
    from these bright objects.}
    {We have selected a sample of 14 HLIRGs observed by
    XMM-\textsl{Newton} (type 1, type 2 AGN and starburst), 5 of which
    are candidates to be Compton-thick objects. This is the first time
    that a systematic study of this type of objects is carried out in
    the X-ray spectral band. Their X-ray spectral properties have been
    correlated with their infrared luminosities, estimated by
    \textsl{IRAS}, \textsl{ISO} and sub-millimeter observations.}
    {The X-ray spectra of HLIRGs present heterogeneous properties. 
    All our X-ray detected HLIRGs (10) have AGN-dominated X-ray
    spectra. The hard X-ray luminosity of 8 of them is consistent with
    a pure AGN contribution, while in the remaining 2 sources both an
    AGN and a starburst seem to contribute to the overall emission. We
    found soft excess emission in 5 sources. In one of them it is
    consistent with a pure starburst origin, while in the other 4
    sources it is consistent with an AGN origin. The observed X-ray
    emission is systematically below the one expected for a standard
    local QSO of the same IR luminosity, suggesting the possible
    presence of absorption in type 2 objects and/or a departure from a
    standard spectral energy distribution of QSO. The
    X-ray-to-IR-luminosity ratio is constant with redshift, indicating
    similar evolutions for the AGN and starburst component, and that
    their respective power sources could be physically related.}
   {}


   \maketitle

\begin{table*}
 \begin{minipage}[!h]{\textwidth}
 \caption{Hyper-luminous galaxies observed by XMM-\textsl{Newton} with redshift less than $\sim2$.}
 \label{tab:sample}
 \centering
 \renewcommand{\footnoterule}{}  
  \begin{tabular}{@{}llclcccc@{}}
  \hline \hline
   Source                       & Type$^a$                    &  \multicolumn{2}{c}{IR-Model$^b$}            &     RA     &  DEC      & $z$ & L$_{FIR}$ \\
                                &                             &         AGN / SB           &         &            &       &    &   [erg s$^{-1}$ cm$^{-2}$] \\   
  \hline
  \object{IRAS 00182-7112}      & \scriptsize{QSO 2  (CT)}    & 0.35 / \textbf{0.65} & RR00 & 00 20 34.7 & -70 55 27 & 0.327   &    46.49 \\ 
  \object{IRAS F00235+1024}     & \scriptsize{NL SB  (CT)}    & 0.5 / 0.5             & F02a & 00 26 06.5 & +10 41 32 & 0.575  & $<$46.76 \\
  \object{IRAS 07380-2342}      & \scriptsize{NL         }    & \textbf{0.6} / 0.4   & F02a & 07 40 09.8 & -23 49 58 & 0.292   &    46.56 \\ 
  \object{IRAS 09104+4109}$^c$  & \scriptsize{QSO 2 (CT), cD} & \textbf{1} / 0       & RR00 & 09 13 45.4 & +40 56 28 & 0.442   & $<$46.42 \\
  \object{PG 1206+459}          & \scriptsize{QSO        }    & \textbf{1} / 0       & RR00 & 12 08 58.0 & +45 40 36 & 1.158   &    47.20 \\
  \object{PG 1247+267}          & \scriptsize{QSO        }    & \textbf{1} / 0       & RR00 & 12 50 05.7 & +26 31 08 & 2.038   &    47.70 \\
  \object{IRAS F12509+3122}     & \scriptsize{QSO        }    & \textbf{0.6} / 0.4   & F02a & 12 53 17.6 & +31 05 50 & 0.780   & $<$46.86 \\  
  \object{IRAS 12514+1027}      & \scriptsize{Sy2    (CT)}    & 0.4 / \textbf{0.6}   & RR00 & 12 54 00.8 & +10 11 12 & 0.3     &    46.18 \\
  \object{IRAS 13279+3401}      & \scriptsize{QSO$^d$    }    & \textbf{0.7} / 0.3   & F02a & 13 30 15.3 & +33 46 29 & 0.36    &    46.58 \\
  \object{IRAS 14026+4341}      & \scriptsize{QSO 1.5$^d$}    & \textbf{0.6} / 0.4   & F02a & 14 04 38.8 & +43 27 07 & 0.323   &    46.26 \\
  \object{IRAS F14218+3845}     & \scriptsize{QSO        }    & 0.2 / \textbf{0.8}   & F02a & 14 23 55.0 & +38 32 14 & 1.21    &    47.80 \\
  \object{IRAS F15307+3252}     & \scriptsize{QSO 2  (CT)}    & \textbf{0.7} / 0.3   & V02  & 15 32 44.0 & +32 42 47 & 0.926   & $<$47.07 \\
  \object{IRAS 16347+7037}      & \scriptsize{QSO        }    & \textbf{0.8} / 0.2   & F02b & 16 34 28.9 & +70 31 33 & 1.334   &    47.42 \\
  \object{IRAS 18216+6418}$^c$  & \scriptsize{QSO 1.2, cD}    & \textbf{0.6} / 0.4   & F02a & 18 21 57.3 & +64 20 36 & 0.297   &    46.49 \\
  \hline
  \end{tabular}
   \footnotetext[1]{NL: narrow-line objects; Sy2: Seyfert 2. 
                    Compton-thick candidates are labeled as
                    CT. Spectral classification from \citet{Rowan00},
                    except IRAS~F00235+1024 \citep{Verma02},
                    IRAS~14026+4321 \citep{Wang06} and IRAS~18216+6418
                    \citep{VeronCetty06}. IRAS~00182-7112 has been
                    classify as type 2 source using the optical
                    emission lines from \citet{Armus89} and the
                    diagnostic diagrams from \citet[ chap. 12]{Osterbrock89}.}
   \footnotetext[2]{Fraction of the IR emission originated in AGN and/or SB. See Sect.~\ref{sec:sample} for details. 
                    Data from: (RR00) \citealt{Rowan00}, (F02a) \citealt{Farrah02submm}, (V02) \citealt{Verma02}, (F02b) \citealt{Farrah02hst}.}
   \footnotetext[3]{Source in cluster.}
   \footnotetext[4]{Not detected in X-rays. We use the optical data to classify this 
                    source as QSO. See Sect.~\ref{sec:sample} for details.}
 \end{minipage}
\end{table*}

\section{Introduction}
\label{sec:intro}
Ultra-luminous Infrared Galaxies (ULIRGs) are a class of galaxies with
bolometric luminosity L$_{IR} \geq 10^{12}$L$_{\odot}$, dominated by
the emission in the infrared (IR) waveband (see \citealt{Sanders96}
for a complete review). They are, together with optical quasars, the
most luminous objects in the Local Universe. ULIRGs are rare in the
Local Universe \citep{Soifer87}, but large numbers are detected
instead in deep-IR surveys, and are a fundamental constituent of the
high redshift galaxy population \citep{Smail97,Genzel00,Fran01}. They
are powered by Active Galactic Nucleus (AGN) and/or starburst (SB)
triggered by mergers of gas-rich spiral galaxies
\citep{Veilleux02}. Optical spectroscopic studies reveal that the
fraction of ULIRGs hosting an AGN increases with increasing IR
luminosity \citep{Veilleux95,Veilleux99}. Furthermore, the fraction of
Seyfert 1 to Seyfert 2 ULIRGs increases with IR luminosity.

It has been proposed that ULIRGs at high redshift could be the
origin of massive elliptical and S0 galaxies
\citep{Fran94,Lilly99,Genzel00}. An important fraction of stars in
present day galaxies would have been formed during these evolutionary
phases.

Observations in the X-ray band are a powerful tool to disentangle the
AGN contribution to the bolometric luminosity from the ULIRGs. X-rays
studies of ULIRGs have confirmed their composite nature (powered by
AGN and/or starburst), with indications for a predominance of the SB
over the AGN phenomenon, even when observed in hard X-rays \citep{Fran03}.

The brightest end of the ULIRG distribution is occupied by the
Hyper-Luminous Infrared Galaxies (HLIRGs, L$_{IR} \geq
10^{13}$L$_{\odot}$). They are among the most luminous objects in the
Universe, although the origin of this luminosity is still
uncertain. This population exhibits extremely high star formation
rates, and most seem to also harbour an AGN
\citep{Rowan00}.

The source and trigger of the emission from HLIRGs have been discussed
since their discovery. Initially the extreme luminosity of HLIRGs was
attributed to gravitational magnification, but \textit{Hubble Space
Telescope} (\textsl{HST}) observations discovered that only a minority
of these galaxies ($\sim$15-20 per cent) have been misclassified owing
to lensing \citep{Farrah02hst}. Currently, there are three main
hypotheses on the nature of these objects:

\begin{description}
\item[a)] They could be simply the high luminosity tail of the ULIRG 
population, and therefore their power sources are probably triggered by 
galaxy mergers. Though \textsl{HST} has revealed some merging 
systems among HLIRGs, there is a significant fraction of them in 
isolated systems \citep{Farrah02hst}.

\item[b)] Assuming that the majority of the rest-frame far infrared (FIR) 
and sub-millimeter (sub-mm) emission comes from star formation
\citep{Verma02,Farrah02submm}, their estimated star formation rates (SFR)
are $>1000M_{\odot}yr^{-1}$, the highest for any object in the
Universe. HLIRGs could be very young galaxies going through their
major episode of star formation \citep{Rowan00}.

\item[c)] They may be a completely new class of objects, where the IR emission is 
originated via some other mechanism: e.g., a transient IR-luminous phase in quasar 
evolution \citep{Farrah02hst}. 
\end{description}

The state-of-the-art X-ray, IR and sub-mm observations suggest that
HLIRGs are powered by dust-enshrouded black hole accretion and violent
star formation \citep{Rowan00,Verma02,Farrah02submm,Wilman03,Iwasawa05}, 
pointing to a mixture of the (a) and (b) hypotheses. Recent
observations of individual HLIRGs with \textsl{Chandra} and
XMM-\textsl{Newton} show that the IR emission of these objects could
be powered by buried quasars through dust re-radiation. The nuclear
source is heavily obscured, reaching the Compton-thick (CT) limit
\citep{Iwasawa01,Wilman03,Iwasawa05}. It has also been suggested that
galaxy mergers in an over-density region may be a necessary condition
for the formation of this class of sources \citep{Iwasawa05}. However
this suggestion is in contradiction with the significant fraction of
isolated HLIRGs found by \citet{Farrah02hst}.

AGN research has taken a central role in the study of galaxy formation
and evolution since the discovery that most local spheroidal galaxy
components (elliptically galaxies and the bulge of spiral galaxies)
contain a super-massive black hole with the mass being proportional to
the stellar mass of the galaxy spheroid \citep{Magorrian98,Mclure02},

These facts are most easily explained if the formation of the
super-massive black hole and the spheroid of galaxies were coeval,
i.e., the black hole was built up by accretion of the same gas that
rapidly formed the stars of the spheroid \citep{Page04, DiMatteo05}.

Additional evidence for the coeval hypothesis is the similarity
between the evolution of cosmic star formation and luminous AGN activity,
since both were much higher in the past up to redshift $\sim2$, than
in the present day \citep{Fran99,Silverman05}. As HLIRGs could
represent the most vigorous stage of galaxy formation, they are unique
laboratories to investigate extremely high stellar formation, and its
connection to super-massive black hole growth.

The main objective of this paper is to determine the relative
contribution of SB and AGN emission to the bolometric luminosity in
HLIRGs, and the interplay between those two phenomena, as well as how
their relative contribution varies with cosmic time.  In our study we
have also compared the properties of HLIRGs to those of ULIRGs.

We have used X-ray observations of HLIRGs from XMM-\textsl{Newton} 
and also public data from the \textsl{Chandra} archive. In
Sect.~\ref{sec:sample} the sample selection is explained. Data
reduction and spectral analysis of each source are described in
Sect.~\ref{sec:xraydata}. Results are discussed in
Sect.~\ref{sec:discuss}. Section \ref{sec:conc} summarizes our
conclusions.

The \textit{Wilkinson Microwave Anisotropy Probe} (\textsl{WMAP})
concordance cosmology has been adopted in the paper: $H_0=70$ km
s$^{-1}$ Mpc$^{-1}$, $\Omega_m=0.27,
\Omega_{\Lambda}=0.73$ \citep{Spergel03}.

\begin{table*}
 \begin{minipage}[!ht]{\textwidth}
 \caption{XMM-\textsl{Newton} observations description.}
 \label{tab:xmmobs}
 \centering
 \renewcommand{\footnoterule}{}  
  \begin{tabular}{@{}lcccccccccll@{}}
  \hline \hline
    Source & \multicolumn{3}{c}{Net. exp. time [ks]} & \multicolumn{3}{c}{Ext.radius [arcsec]}  & \multicolumn{3}{c}{Source counts$^a$} & Obs. date & Filter \\
                        &  PN & MOS1 & MOS2 &                      PN & MOS1 & MOS2                       &  PN & MOS1 & MOS2 &          &                 \\
  \hline
   IRAS 00182-7112      &  9.1 & -    & -    & 19 & - & -   &   $134\pm15$  & -           & -           & 2003-04-17     & Thin   \\
   IRAS F00235+1024     & 14.4 &  -   & -    & 35 & -  & -  &     $<30$     & -           & -           & 2001-01-10     & Thin   \\
   IRAS 07380-2342      &  4.7 & 11.4 & 11.6 & 30 & 40 & 40 &     $<45$     & -           & -           & 2005-10-13$^c$ & Medium \\
   IRAS 09104+4109$^b$  &  9.2 & 12.6 & 12.5 & 20 & 20 & 20 &  $5544\pm75$  & $2245\pm47$ & $2332\pm48$ & 2003-04-27     & Medium \\
   PG 1206+459          &  5.9 &  6.9 &  6.9 & 26 & 32 & 29 &   $731\pm32$  &  $188\pm16$ &  $193\pm16$ & 2002-05-11     & Thin   \\
   PG 1247+267          & 19.4 & 25.5 & 26.6 & 35 & 35 & 35 &  $5132\pm74$  & $1646\pm42$ & $1694\pm43$ & 2003-06-18     & Medium \\
   IRAS F12509+3122     & 11.9 & 14.6 & 14.1 & 25 & 25 & 25 &   $508\pm24$  &  $156\pm13$ &  $139\pm13$ & 2005-12-11$^c$ & Thin   \\
   IRAS 12514+1027      & 16.7 & 20.0 & 18.2 & 22 & 40 & 40 &   $105\pm22$  &   $28\pm13$ &    $7\pm12$ & 2001-12-28     & Thin   \\
   IRAS 13279+3401      & 24.6 &  -   &   -  & 25 & -  & -  &    $<36$      &      -      &      -      & 2006-01-18$^c$ & Medium \\
   IRAS 14026+4341      &  -   &  6.6 &  5.6 & -  & 35 & 35 &       -       &     $<30$   &     $<27$   & 2005-11-26$^c$ & Thin   \\
   IRAS F14218+3845     & 11.5 & 15.2 & 15.1 & 30 & 34 & 31 &   $550\pm29$  &  $185\pm17$ &  $176\pm16$ & 2003-08-01     & Medium \\
                        &  2.3 &  7.3 &  7.0 & 30 & 30 & 30 &    $91\pm14$  &  $109\pm20$ &   $79\pm18$ & 2005-06-07$^c$ & Medium \\
   IRAS F15307+3252     &  9.4 & 11.5 & 12.0 & 22 & 40 & 40 &    $97\pm21$  &   $30\pm12$ &   $42\pm13$ & 2002-07-30     & Medium \\
   IRAS 16347+7037      & 12.9 & 15.6 & 15.8 & 30 & 56 & 48 & $11172\pm106$ & $3546\pm67$ & $3785\pm66$ & 2002-11-23     & Medium \\
   IRAS 18216+6418$^b$  &  0.5 &  2.9 &  3.3 & 20 & 20 & 20 &       -       & $7402\pm86$ & $8481\pm92$ & 2002-10-16     & Thin   \\
  \hline
  \end{tabular}
 \footnotetext[1]{Total source counts in the 0.2-10 keV band.}
 \footnotetext[2]{\textsl{Chandra} data are also used in the study of this source. 
                  See Section \ref{sec:clsubtraction} for details.}
 \footnotetext[3]{Data from OBS-ID 030536.}
 \end{minipage}
\end{table*}

\section{The HLIRG sample}
\label{sec:sample}
Our sample of HLIRGs has been selected from the \citet{Rowan00} sample
of 45 HLIRGs. The sources in this mother sample can be classified in
four sub-samples: (1) objects found from direct optical follow-up of
60~$\mu m$ or 850~$\mu m$ surveys; (2) sources found from comparison
of known quasar and radio galaxy lists with 60~$\mu m$ catalogues, or
using IR color selection (biased to AGN); (3) sources selected from
sub-mm observations of known high-redshift AGN; and (4) known luminous
IR galaxies not included in the previous subsamples, but satisfying
L$_{IR}>10^{13}$L$_\odot$. The first one is a flux limited sample,
unbiased towards AGN; the sources in the second and third sub-samples
have been selected in order to host an AGN, and therefore suffer from
selection effects.

We have chosen all the HLIRGs with public data in the
XMM-\textsl{Newton} Science Archive (XSA), as of December 2004. We
included also observations of five sources from OBS-ID 030536 by our
group (see Table~\ref{tab:xmmobs}). Then we constrained the resulting
sample to those sources with redshift less than $\sim$2, to prevent a
strong bias due to the presence of high-$z$ QSOs. In our final sample
there are 8 sources which are included in the first \citet{Rowan00}
subsample, 4 sources are in the second subsample, 1 source is in the
third and 1 is in the fourth one. Most of our sources are selected
from subsamples which are in principle not biased in favour of
AGN. However, selecting the sample by using the availability of
XMM-\textsl{Newton} data probably introduces a selection
effect in favour of the presence of an AGN. Moreover,
estimating the completeness level of this sample is
difficult, since it is not flux limited.  From IR and sub-mm unbiased
surveys, \citet{Rowan00} estimates that the number of HLIRGs brighter
than 200~mJy at 60~$\mu m$ over the whole sky is 100-200. 14 of them
are included in our sample, which is the largest sample of HLIRGs
studied in X-rays.

Table~\ref{tab:sample} describes our sample. Column 2 shows the
optical spectral classification as derived from the literature: twelve
sources in our sample present AGN characteristics. Eight of them are
classified as `type 1', and four of them as `type 2'. We have
classified QSO instead of Seyfert to those objects with intrinsic
2-10~keV luminosity $>10^{44}~\mathrm{erg~s}^{-1}$ (see
Table~\ref{tab:xraydata}).  A couple of sources have been classified
from the literature as ``Narrow line'' (NL) sources, i.e. which show
narrow forbidden emission lines typical of HII regions. All `type 2'
and one NL-SB galaxy are CT candidates.

The analysis of the IR Spectral Energy Distribution (SED) of 
our sources has revealed that the SED can be modeled by a combination
of an AGN and a SB components \citep{Rowan00,Verma02,Farrah02submm}. In
Table~\ref{tab:sample}, column 3 we report the fraction of the AGN and
SB component needed to fit the SED. Three objects are completely
dominated by the AGN component, while in other three the SB component
is dominant.

In order to compare the properties of HLIRGs with other similar
classes of objects, we have included in our study two samples. We
chose a sample of 10 ULIRGs studied in X-rays by XMM-\textsl{Newton}
\citep{Fran03}. The sample is flux-limited at 60~$\mu m$ and complete
to $S_{60~\mu m} \ge 5.4$~Jy. In addition, we have selected all the
HLIRGs (six) from the \citet{Stevens05} sample of high redshift X-ray
Compton-thin absorbed QSO. These sources have been observed in X-rays
by \textsl{ROSAT} \citep{Page01} and by XMM-\textsl{Newton}
\citep{Page07}, and by SCUBA in the sub-mm band \citep{Page04}.

\begin{figure*}[!ht]
   \begin{center} 
    \mbox{ 
     \subfigure[IRAS 00182-7112: PN]{
      \includegraphics[angle=-90,width=.5\linewidth]{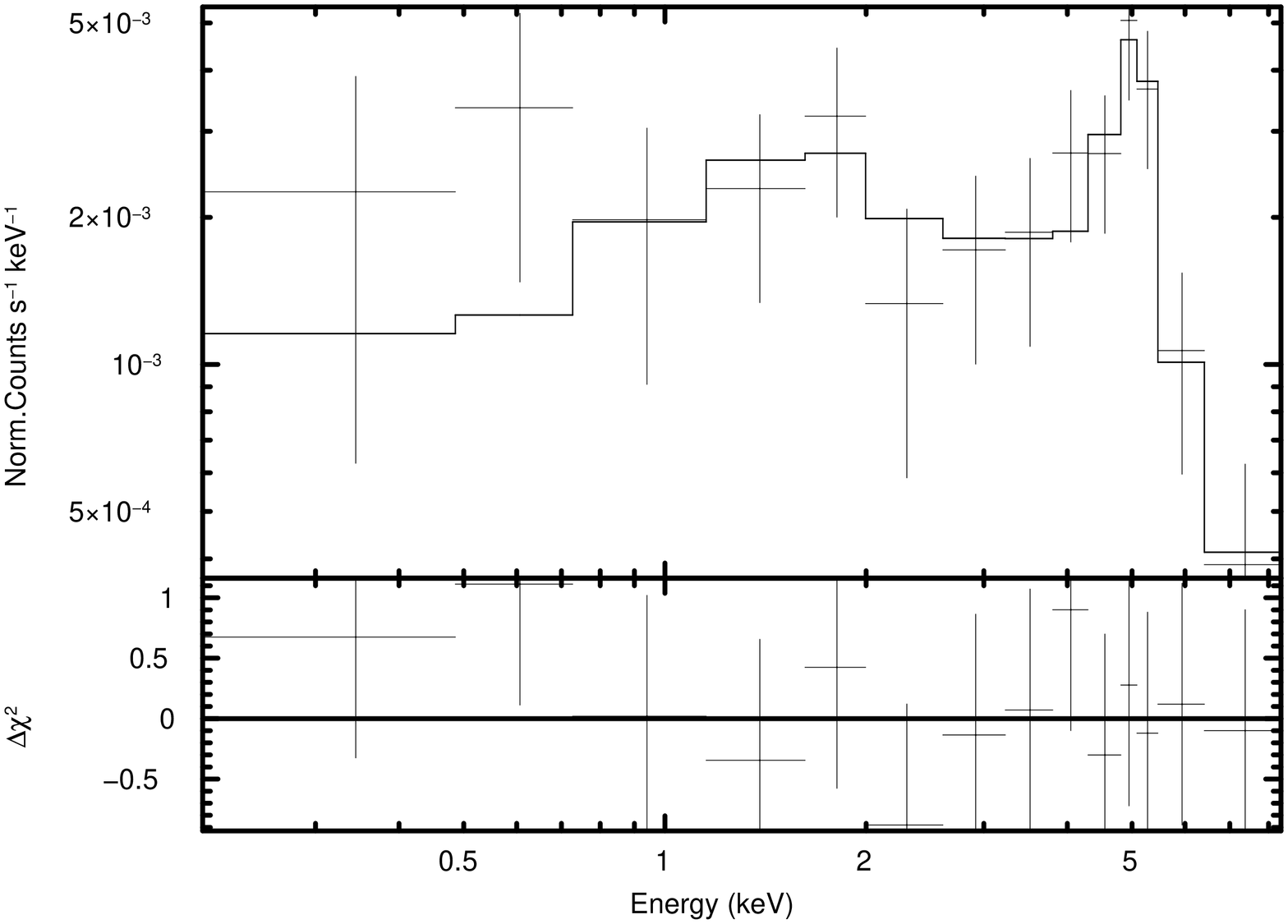}
      \label{fig:iras00182} 
     } 
    \quad 
     \subfigure[IRAS 09104+4109: PN, MOS1, MOS2]{
      \includegraphics[angle=-90,width=.5\linewidth]{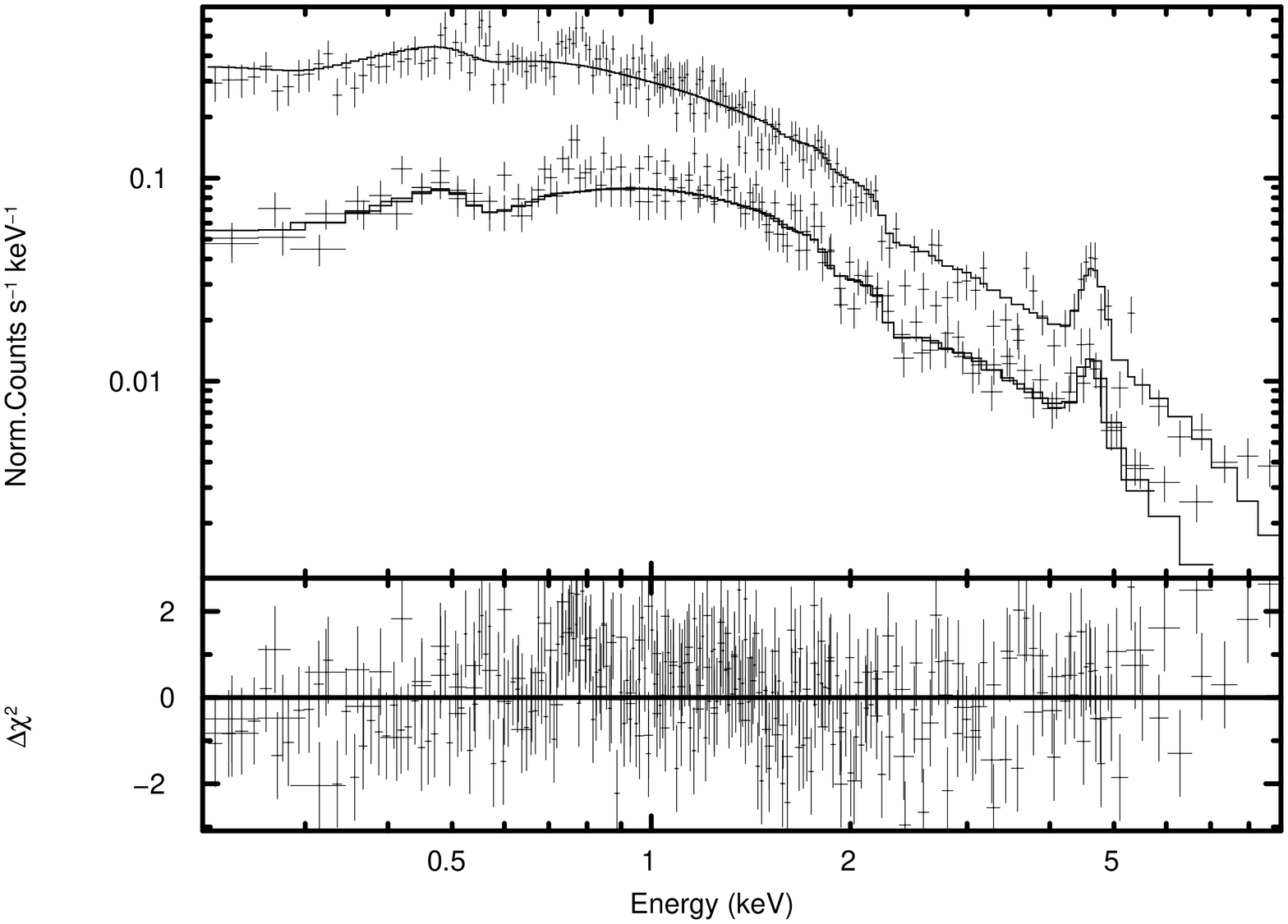}
      \label{fig:iras09104} 
     } 
    } 
    \mbox{ 
     \subfigure[PG 1206+459: PN, MOS1, MOS2]{
      \includegraphics[angle=-90,width=.5\linewidth]{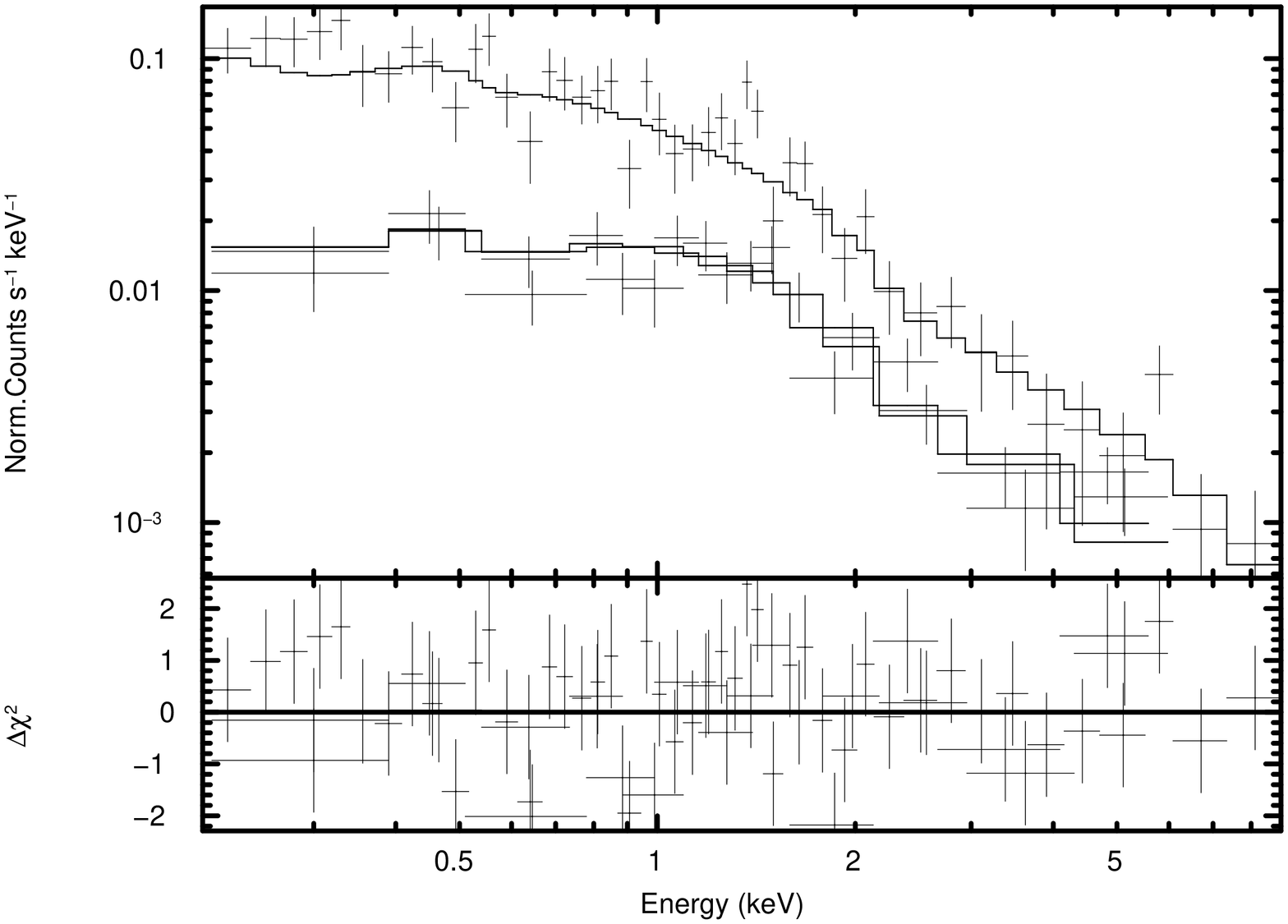}
      \label{fig:pg1206} 
     } 
    \quad 
     \subfigure[PG 1247+267: PN, MOS1, MOS2]{ 
      \includegraphics[angle=-90,width=.5\linewidth]{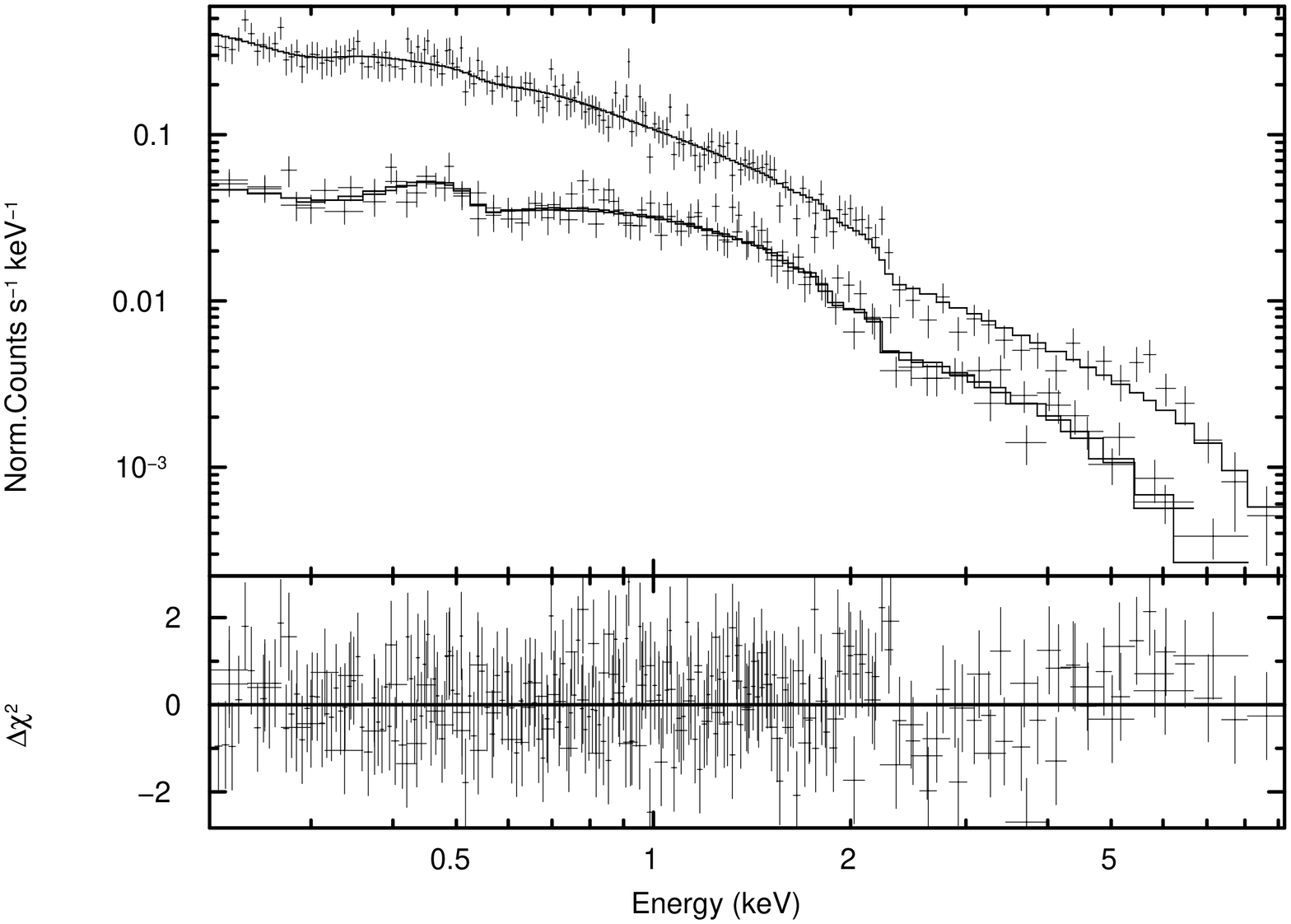}
      \label{fig:pg1247} 
     } 
    } 
    \mbox{ 
     \subfigure[IRAS F12509+3122: PN, MOS1, MOS2]{
      \includegraphics[angle=-90,width=.5\linewidth]{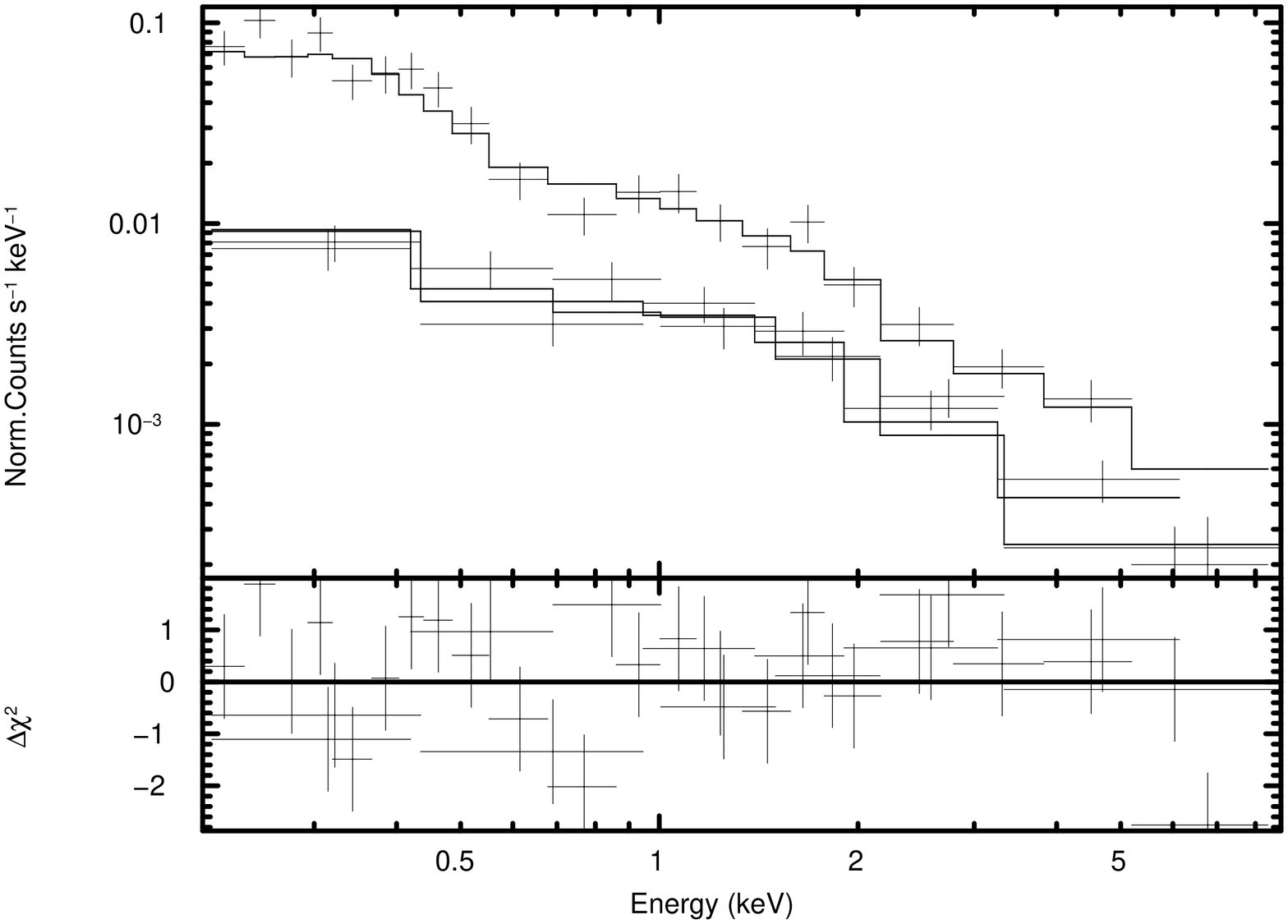}
      \label{fig:pgF12509} 
     } 
    \quad 
     \subfigure[IRAS 12514+1027: PN+MOS1+MOS2]{
      \includegraphics[angle=-90,width=.5\linewidth]{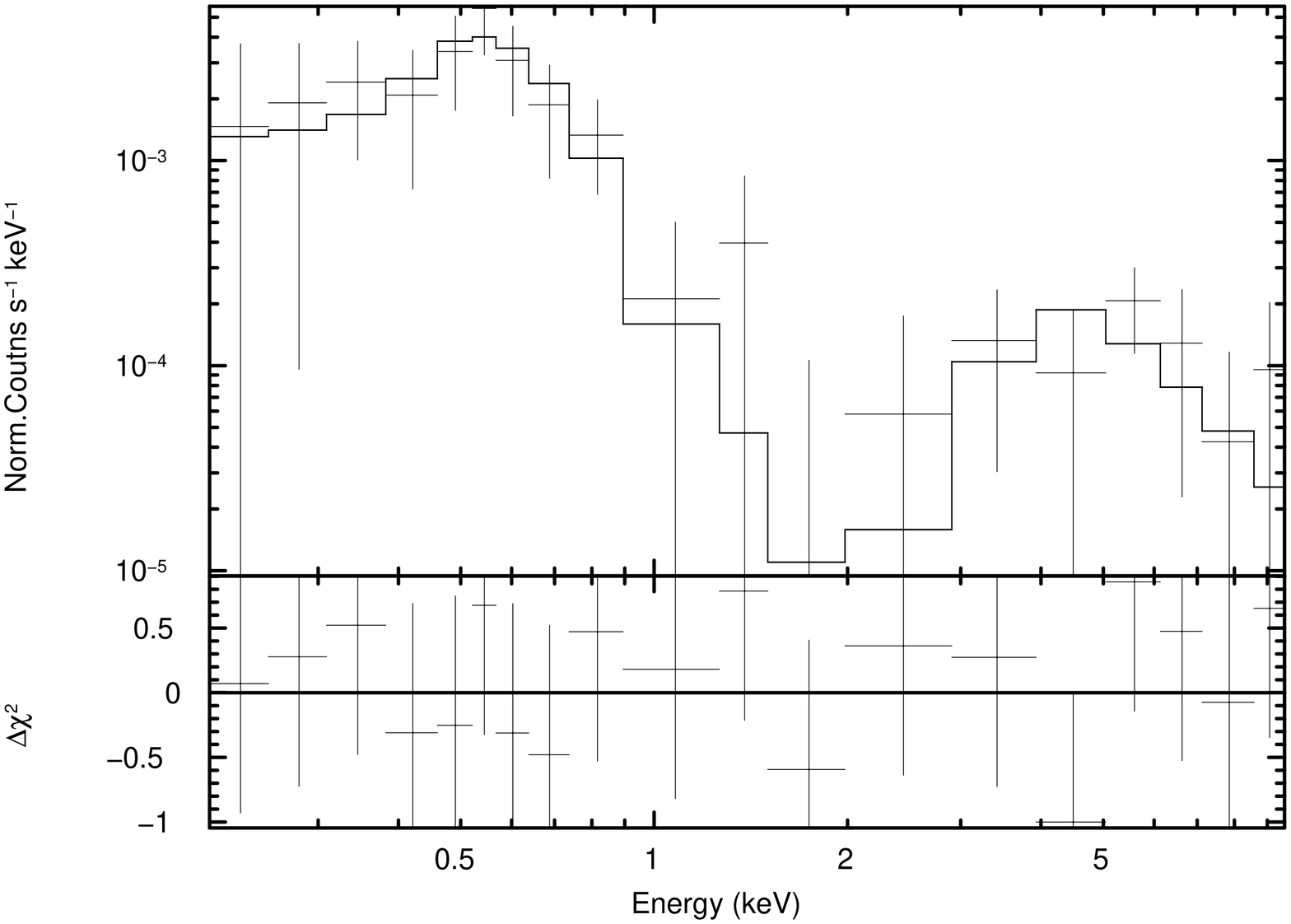}
      \label{fig:iras12514} 
     } 
    } 
   \end{center}
 \caption{XMM-\textsl{Newton} X-ray spectra and residuals of the detected sources 
          from the HLIRGs sample. The solid line is the best fit model. Continues 
          on next figure.}
 \label{fig:specplots}
\end{figure*}

\addtocounter{figure}{-1}

\begin{figure*}[!ht]
 \addtocounter{subfigure}{6}
   \begin{center}
      \mbox{
	\subfigure[IRAS F14218+3845: PN+MOS1+MOS2 (two observations)]{
	   \includegraphics[angle=-90,width=.5\linewidth]{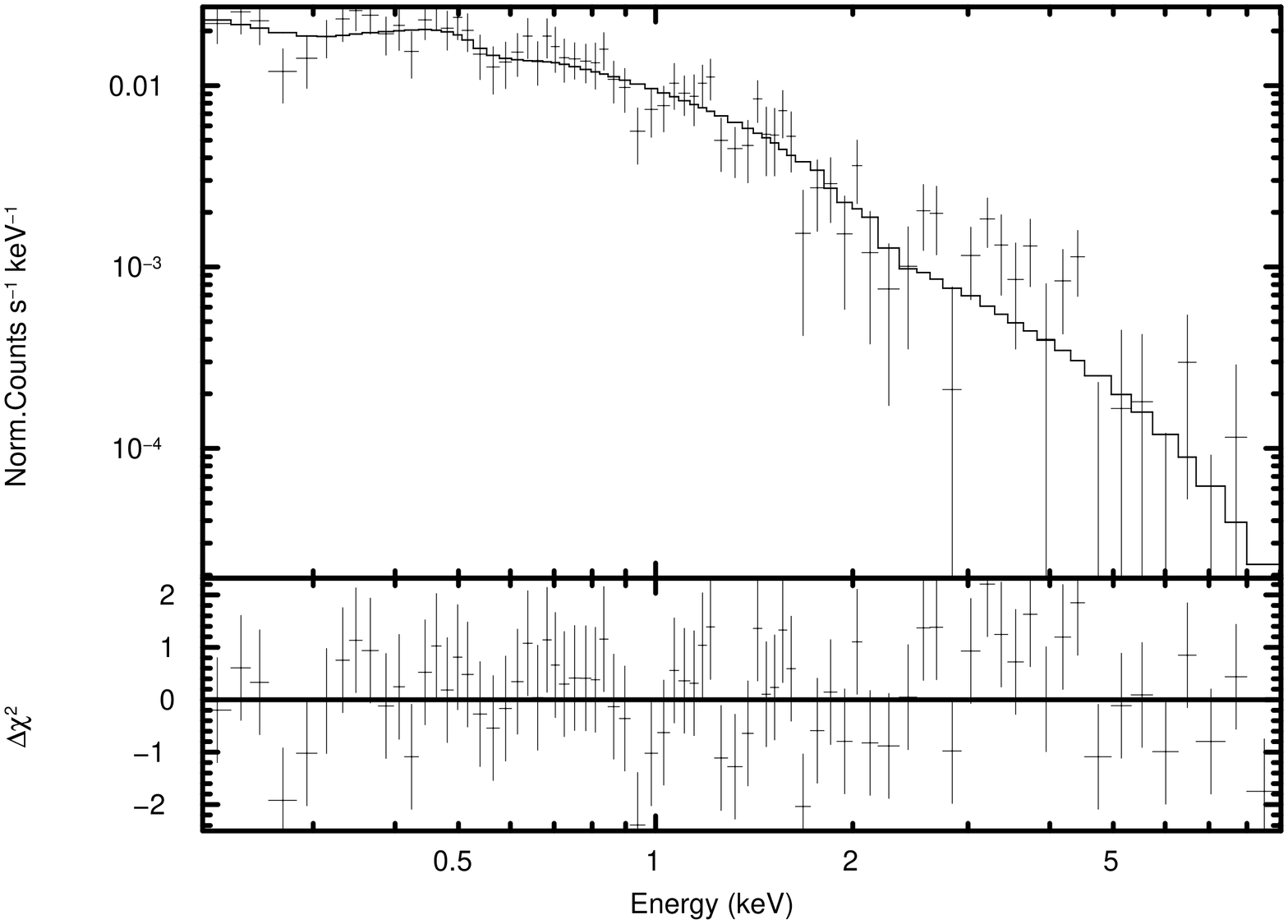}
	   \label{fig:irasF14218}
	} \quad
	\subfigure[IRAS F15307+3252: PN+MOS1+MOS2]{
	   \includegraphics[angle=-90,width=.5\linewidth]{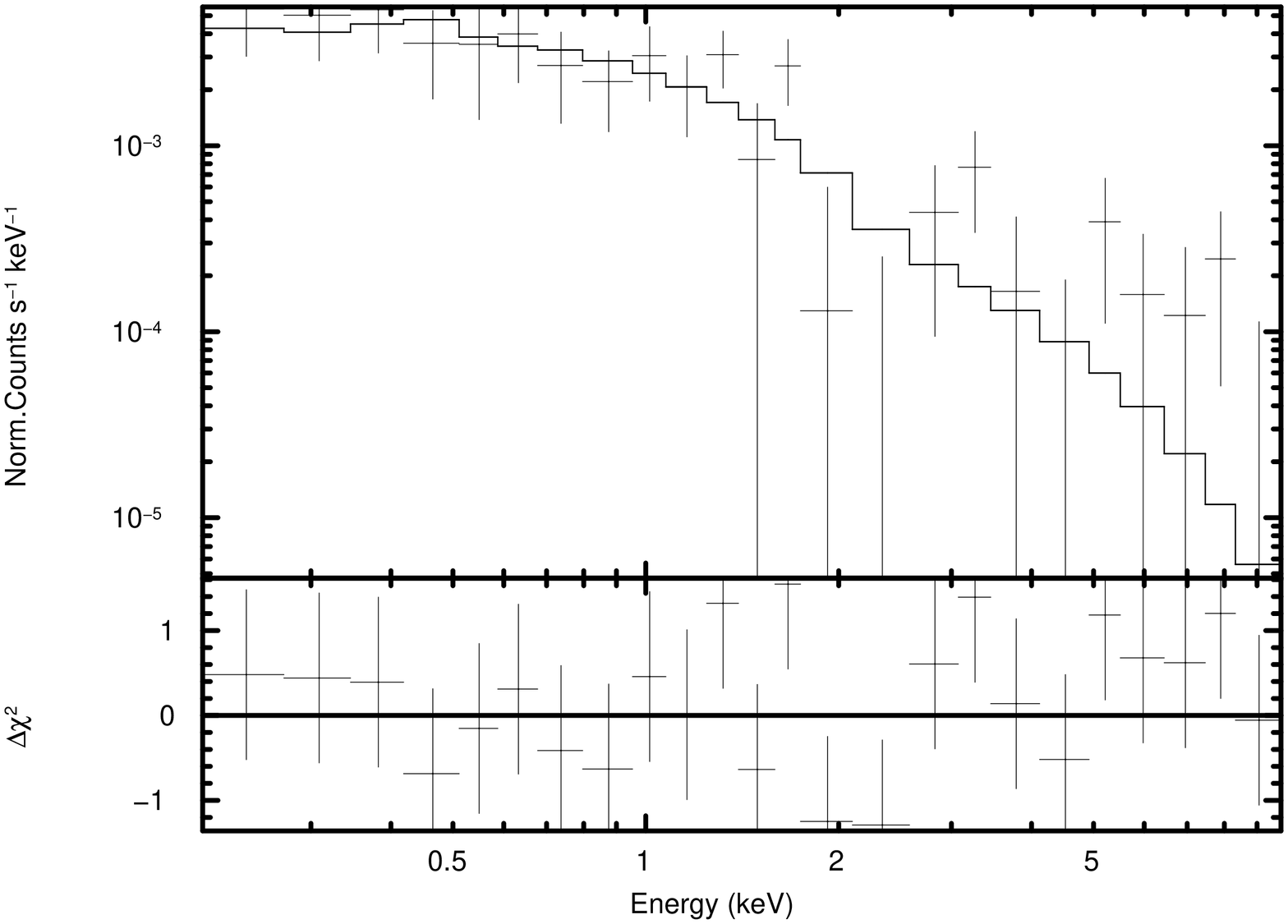}
	   \label{fig:irasF15307}
	}
      }
      \mbox{
	\subfigure[IRAS 16347+7037: PN, MOS1, MOS2]{
	   \includegraphics[angle=-90,width=.5\linewidth]{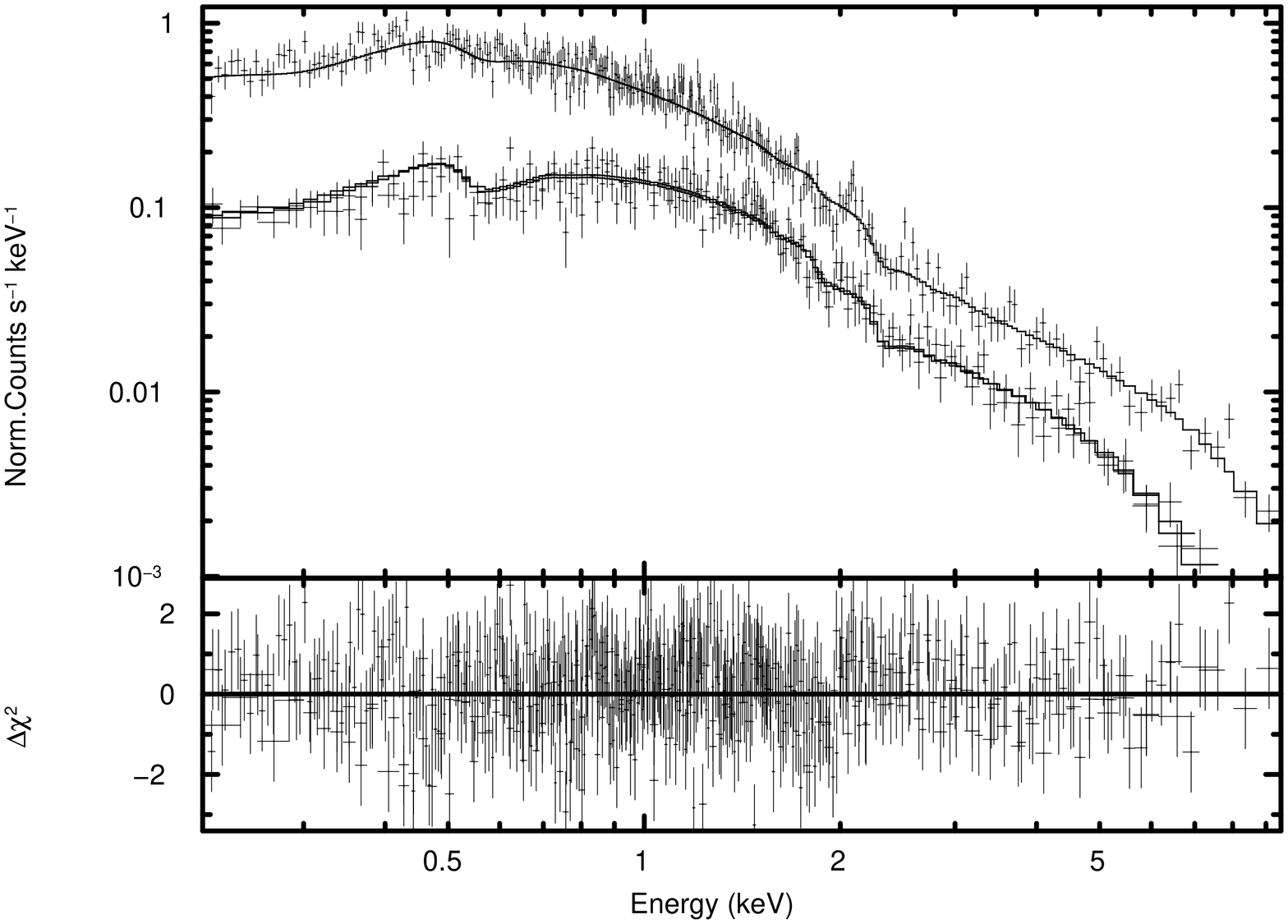}
	   \label{fig:iras16347}
	} \quad
	\subfigure[IRAS 18216+6418: MOS1, MOS2]{
	   \includegraphics[angle=-90,width=.5\linewidth]{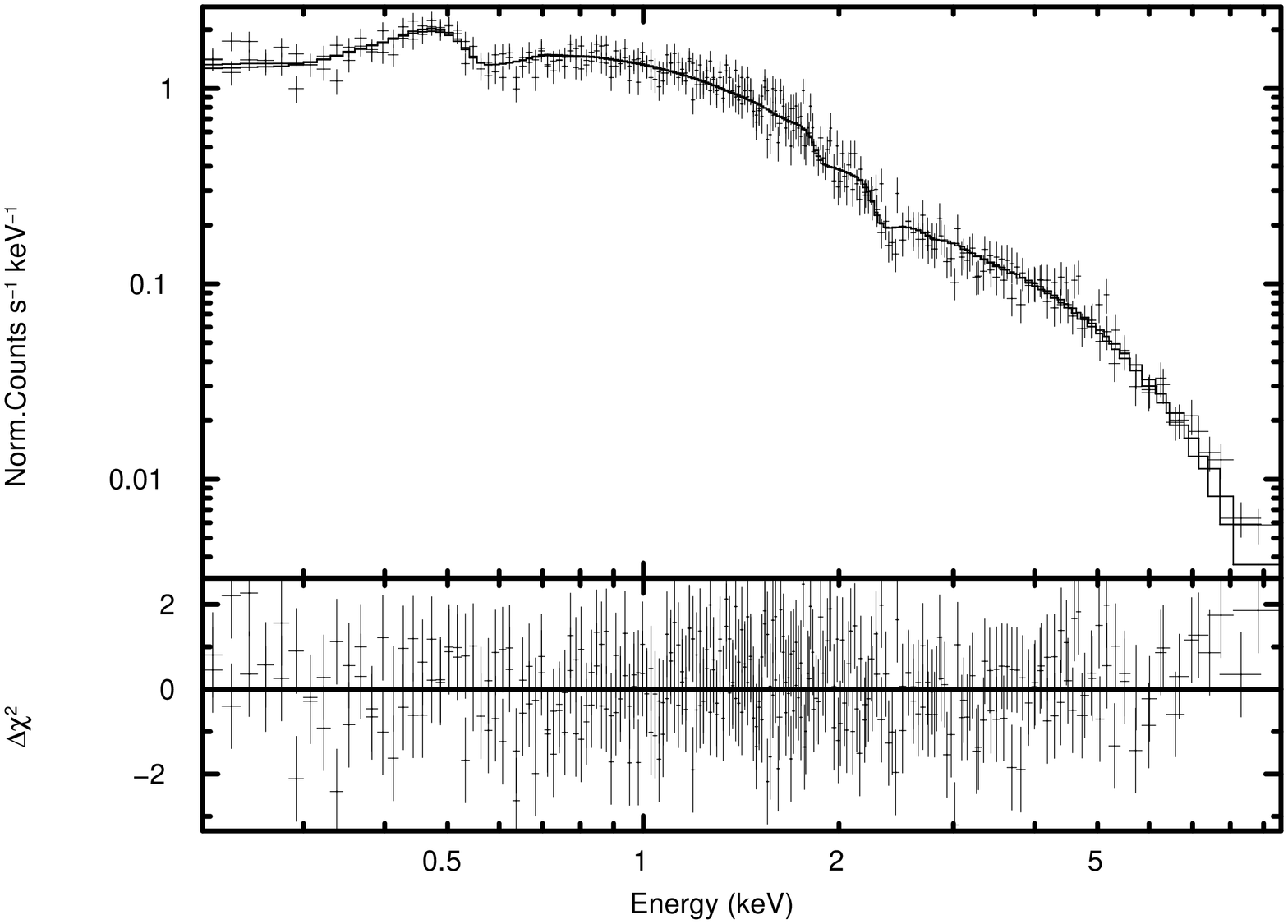}
	   \label{fig:iras18216}
	}
      }
   \end{center}
\caption{Continue}
\label{fig:specplots:b}
\end{figure*}

\section{X-ray data reduction and analysis}
\label{sec:xraydata}

\subsection{Data reduction}
\label{sec:reduction}
Table \ref{tab:xmmobs} presents the most relevant information about
the XMM-\textsl{Newton} observations. The data have been processed using
the Science Analysis System (SAS) version 6.1.0, and have been
analyzed using the standard software packages (FTOOLS and XSPEC)
included in HEAsoft 5.3.1.

We reprocessed the EPIC PN and MOS Observation Data Files (ODFs) to
obtain new calibrated and concatenated event lists, using the SAS
tasks \texttt{EMPROC} and \texttt{EPPROC}, including the latest
calibration files at the time of reprocessing.

The new event files were filtered to avoid intervals of flaring
particle background, and only events corresponding to pattern 0-12 for
MOS and 0-4 for PN were used \citep{xmmhandbook}. The events with
energy above 12~keV and below 0.2~keV were also filtered out. The
source spectra were extracted from circular regions, whose radius was
chosen in each case to optimize the signal-to-noise ratio (S/N), and
to avoid the CCD gaps. The background spectra were taken in circular
source-free regions near the object, also avoiding CCD gaps. We
generated our own redistribution matrices and ancillary files
(correction for the effective area) using the SAS tasks
\texttt{RMFGEN} and \texttt{ARFGEN}.

XMM-\textsl{Newton} has detected 10 out of 14 sources ($\sim 70\%$)
with different S/N quality. In those cases where the S/N ratio was
poor, the MOS and PN spectra were co-added \citep{Page03coadd}. All
spectra were rebinned to have $\ge25$ counts per energy channel,
except IRAS 00182-7112 ($\ge15$ counts) and PG 1206+459 ($\ge20$
counts). The resulting EPIC spectra (see Fig.~\ref{fig:specplots})
reveal heterogeneous spectral properties for these objects (see Table
\ref{tab:xraydata} and Sects.~\ref{sec:analysis},\ref{sec:xdsources}).

\subsection{Non-detected sources}
\label{sec:upperlimits}
We have estimated upper limits to the luminosity of the sources not
detected by XMM-\textsl{Newton}. We estimated the count rate which
would correspond to 3$\sigma$ fluctuations of the background in a
circular region of the PN-EPIC images, centered in the source
coordinates. To convert between count rate and physical units a simple
model was chosen: a power law with $\Gamma=2$ and Galactic absorption.

\subsection{Cluster emission subtraction}
\label{sec:clsubtraction}
Two sources of our sample, IRAS 09104+4109 and IRAS 18216+6418, reside
in clusters. They present soft extended emission from the
intra-cluster medium (ICM). To take into account this residual
foreground in the subsequent spectral analysis, we added an XSPEC
thermal component to the spectral model. Two parameters characterize
this model: temperature and normalization.

We estimated the temperature of the cluster extracting an X-ray
spectrum in an annular region around the source and fitting it with a
thermal bremsstrahlung model. The normalization was obtained
re-normalizing the flux from the annulus to that from the source
circular region.  To this end, we integrated the X-ray surface
brightness profile of the cluster over both regions\footnote{The
background spectrum of the annular region was extracted in a circular
region free of sources away from the cluster emission. We used the
XMM-\textsl{Newton} ``blank fields'' \citep{xmmblankfields} to extract
the background spectrum of the source.}.

We determined the brightness profile using public \textsl{Chandra}
data (see Table~\ref{tab:chandraobs}). Assuming an isothermal ICM, the
radial profile of a cluster can be well fitted by a $\beta$-model
\citep{Sarazin86,Mushotzky04}. This profile was then convolved with
the XMM-\textsl{Newton} PSF, before integrating it over the regions of
interest.

\begin{table}
 \begin{minipage}[b]{\columnwidth}
 \caption{\textsl{Chandra} observations description.}
 \label{tab:chandraobs}
 \centering
 \renewcommand{\footnoterule}{}  
  \begin{tabular}{@{}lcccc@{}}
  \hline \hline
        Source                   & Instrument & Grating & Exp. time. & Obs. date  \\
                                 &            &         &   [ks]     &            \\ 
  \hline
   \scriptsize{IRAS 09104+4109}  & ACIS-S     &  None   &    9.17    & 1999-11-03 \\ 
   \scriptsize{IRAS 18216+6418}  & ACIS-S     &  LETG   &  171.82    & 2001-01-18 \\   	
  \hline
  \end{tabular}
 \end{minipage}
\end{table}

\subsection{Spectral analysis}
\label{sec:analysis}
Our aim is to estimate the AGN and SB contribution to the total X-ray
emission. The general model of the X-ray spectrum emitted by an AGN
has typically four components: an underlying absorbed power
law\footnote{$\frac{dN}{d\varepsilon dt}\propto\varepsilon^{-\Gamma}$,
where $N$ are the number of counts and $\varepsilon$ is the energy of
the counts.}, a reflection component, an iron K$_\alpha$ emission line
and a soft excess above the power law at energies below $\sim1$~keV.

The typical power law photon index for AGN is $\Gamma\simeq1.5-2.1$
\citep{Nandra94,Reeves00,Mateos05}. The soft excess component, common
in type 1 AGN \citep{Reeves00,Piconcelli05}, can be fitted with a
thermal model, with an expected temperature remarkably constant around
0.1-0.3~keV \citep{Piconcelli05,Gierlinski06}. The ratio between the
luminosity of the soft excess component and the power law component in
the soft band (0.5-2~keV) is $\sim0.15-1.45$ \citep{Piconcelli05}. The
X-ray-to-bolometric luminosity ratio in the hard band (2-10~keV) for
an AGN is $\sim0.015$ (\citealt{Elvis94,Risaliti04}, see
Sect.~\ref{sec:discuss}).

Several models have been proposed for the origin of the observed soft
excess in AGN, such as a relativistically blurred photoionized disc
reflection \citep{Ross93,Crummy06} or ionized absorption in a wind
from the inner disc \citep{Gierlinski04,Gierlinski06}. It has also
been shown that the analysis of the high resolution
XMM-\textsl{Newton} RGS spectra is important to understand the nature
of the soft excess emission \citep{Bianchi06}. We have only used
thermal models in this paper for simplicity, since more detailed
models are not warranted by the quality and low spectral resolution of
the data. A deeper analysis of the soft excess emission of these
sources will be presented elsewhere.

The SB emission can be modeled with a power law with 
a typical photon index $\Gamma\simeq1.0-1.4$ \citep{White83,Dahlem98,Persic02},
or with a thermal model whit temperature $0.5-1.0$~keV
\citep{Iwasawa99,Fran03}. The soft X-ray-to-bolometric luminosity
ratio for a SB is $\sim10^{-4}$, as found by \citet{Iwasawa99} for a
sample of four prototype powerful FIR SB galaxies. We can use this to
estimate the bolometric luminosity of the sources.

All sources were analyzed using the following scheme. First we fitted
the data with a power law model plus Galactic absorption\footnote{All
models referred in this paper include a multiplicative Galactic
absorption component fixed at the Galactic N$_H$ value from
\citet{nH}.} (see Table \ref{tab:xraydata}, column 3). We added
intrinsic absorption (XSPEC \texttt{zpha} model) where it was
statistically significant\footnote{To this end we used the F-test,
accepting additional spectral components only when they improved the
fit with a significance $\ge3\sigma$.}.

We compared this model with a power law reflected from neutral
material (modeled with the XSPEC \texttt{pexrav}\footnote{We kept
fixed all the parameters of the \texttt{pexrav} model to the standard
values except the photon index of the incident power law, the
reflection scaling factor and the normalization.},
\citealt{pexrav}). When two models had the same number of parameters
(and hence the F-test is not useful), the fit with the lowest $\chi^2$
was taken as our baseline model. However, when the $\chi^2$ from two
models were comparable, we adopted as our baseline fit the one with
less uncertainty in the determination of the model parameters.

Then, we checked for any significant additional component: iron
emission line (modeled with \texttt{zgauss} at $\sim6.4$~keV) and/or
soft excess. We fitted the latter with different thermal models
(blackbody: \texttt{zbbody}; bremsstrahlung: \texttt{zbremss},
\citealt{zbremss}; bremsstrahlung with emission lines: \texttt{mekal},
\citealt{mekal1,mekal2}). We introduced this thermal component to
parameterize the starburst emission and/or the AGN soft excess
emission.

In those cases where $\Gamma$ was out of the range expected for AGN or
SB, this parameter was fixed to 2, and we tried to fit again the
spectrum with an absorbed power law and a reflection model
(\texttt{pexrav}). In either case, additional intrinsic absorption and
thermal emission models were also added (if needed) to improve the
fit.

The best fit model for each source is given in
Table~\ref{tab:xraydata}. We have calculated the luminosity for each
component in the hard and soft X-ray bands, corrected by Galactic and
intrinsic absorption. A more detailed description of the analysis and
the results for each source is presented in Section
\ref{sec:xdsources}.

We have estimated upper limits (see Table~\ref{tab:xraydata}, column
11) for a thermal component in the X-ray spectra of those sources 
where it was not significant. To this end, we fixed all parameters to
their best fit values. We added a thermal component (\texttt{zbremss})
with a fixed temperature, $kT=0.6$~keV (the mean temperature of the
ULIRGs thermal component from \citealt{Fran03}), and we
calculated the 2$\sigma$ confidence interval for the normalization
parameter, which was then used to estimate the upper limit to the
luminosity.

\begin{table*}
 \begin{minipage}[t]{\linewidth}
 \caption{XMM-\textsl{Newton} spectral analysis results. Fluxes and luminosities in CGS units. 
                      Errors quoted are 90\% confidence level for one interesting parameter throughout this paper.}
 \label{tab:xraydata}
 \centering
 \renewcommand{\footnoterule}{}  
    \begin{tabular}[h]{llccccccrrrrr}
      \hline \hline
      Source                        & Model~$^a$                       & $N_H$~$^b$ & $\Gamma$               & $kT$                   & EW                      & $\chi^2/\nu$ 
                                                                                                                                                                          & $\Delta\chi^2$ 
                                                                                                                                                                                 & $\log S_X$~$^c$ 
                                                                                                                                                                                            & \multicolumn{4}{c}{$\log L$~$^d$}\\
      \cline{10-13} 	
                       &                                               &            &                        & (keV)                  & (keV)                   &         &      &          & $L_{PL}^{SX}$ & $L_{TH}^{SX}$ & $L_{PL}^{HX}$ & $L_{TH}^{HX}$ \\
      \hline
      \scriptsize{IRAS 00182-7112}  & \scriptsize{D + E}               &  4.24      & 2, $E_c=9^{+21}_{-4}$  & -                      & $0.8\pm0.6$             & 5.0/9   & 10.7 & -12.8    &     44.9 & $<$41.9 &     44.8 & $<$40.7 \\
      \scriptsize{IRAS F00235+1024} & -                                &  5.07      & 2                      & -                      & -                       &    -    &  -   & $<$-14.5 &  $<$42.2 &       - &  $<$42.4 &       - \\
      \scriptsize{IRAS 07380-2342}  & -                                &  64.3      & 2                      &                        & -                       &    -    &  -   & $<$-13.7 &  $<$41.7 &       - &  $<$42.5 &       - \\
      \scriptsize{IRAS 09104+4109}  & \scriptsize{(B)$^e$ + A + E}     &  1.82      & $1.62\pm0.07$          & -                      & $0.38^{+0.10}_{-0.11}$  & 374/324 & 42   & -11.7    &     44.2 & $<$43.0 &     44.5 & $<$41.8 \\   
       \scriptsize{+\textsl{Beppo}SAX:} & \scriptsize{(B)$^e$ + B + D + E} &  $''$ & $1.2^{+0.3}_{-0.2}$     & $3.1^{+0.4}_{-0.3}$    & $0.2^{+3.4}_{-0.1}$     & 330/326 & 53   &  $''$    &     44.7 &    44.2 &     45.3 &    44.2 \\
      \scriptsize{PG 1206+459}      & \scriptsize{A}                   &  1.31      & $1.7\pm0.9$            & -                      & -                       &  72/67  &  -   & -12.5    &     45.8 & $<$44.0 &     45.1 & $<$42.8 \\
      \scriptsize{PG 1247+267}      & \scriptsize{A + C}               &  0.90      & $1.98^{+0.05}_{-0.07}$ & $0.49^{+0.23}_{-0.17}$ & -                       & 235/281 & 30   & -12.3    &     45.9 &    45.5 &     45.9 &    44.0 \\                 
      \scriptsize{IRAS F12509+3122} & \scriptsize{A + B}               &  1.24      & $1.38^{+0.11}_{-0.12}$ & $0.21\pm0.03$          & -                       &  38/30  & 80   & -12.9    &     43.8 &    43.8 &     44.3 &    40.2 \\
      \scriptsize{IRAS 12514+1027}  & \scriptsize{B + F $\times$ A}$^f$ & 1.67      & 2                      & $0.35^{+0.17}_{-0.07}$ & -                       &   5/15  & 15   & -14.2    &     43.2 &    42.2 &     43.3 &    39.7 \\
      \scriptsize{IRAS 13279+3401}  & -                                &  0.99      & 2                      & -                      & -                       &    -    &  -   & $<$-14.2 &  $<$42.1 &       - &  $<$42.2 &       - \\
      \scriptsize{IRAS 14026+4341}  & -                                &  1.19      & 2                      & -                      & -                       &    -    &  -   & $<$-13.7 &  $<$42.5 &       - &  $<$42.6 &       - \\
      \scriptsize{IRAS F14218+3845} & \scriptsize{A}                   &  0.93      & $2.24\pm0.12$          & -                      & -                       &  70/75  &  -   & -13.1    &     44.7 & $<$43.8 &     44.6 & $<$42.5 \\
      \scriptsize{IRAS F15307+3252} & \scriptsize{A}                   &  2.03      & $2.1\pm0.4$            & -                      & -                       &  16/22  &  -   & -13.6    & 45.4$^g$ & $<$43.1 & 45.5$^g$ & $<$41.9 \\
      \scriptsize{IRAS 16347+7037}  & \scriptsize{C + A}               &  4.48      & $1.77\pm0.13$          & $1.53\pm0.18$          & -                       & 561/547 & 26   & -11.7    &     45.8 &    45.7 &     46.0 &    45.4 \\
      \scriptsize{IRAS 18216+6418}  & \scriptsize{(C)$^e$ + C + A}     &  4.04      & $1.57^{+0.10}_{-0.12}$ & $0.49^{+0.09}_{-0.08}$ & -                       & 329/332 & 151  & -10.7    &     45.2 &    45.1 &     45.6 &    43.6 \\ 
      \hline 
    \end{tabular}
   \footnotetext[1]{XSPEC models: A: \texttt{power law}, B: \texttt{mekal}, C: \texttt{zbremss}, 
                    D: \texttt{pexrav}, E: \texttt{zgauss} (emission line), F: \texttt{zpha} (intrinsic absorption).}
   \footnotetext[2]{Galactic neutral hydrogen column density (in units of $10^{20}$ atoms cm$^{-2}$), from \citet{nH}.}	
   \footnotetext[3]{Observed frame 0.5-10~keV flux.}
   \footnotetext[4]{Rest frame intrinsic X-ray luminosity. \\
                    SX: Soft X-ray (0.5-2 keV) band; HX: Hard X-ray (2-10 keV) band; 
                    PL: Power law component; TH: Thermal (soft excess) component.}
   \footnotetext[5]{This component models the cluster contribution.}
   \footnotetext[6]{Intrinsic $N_H=4^{+20}_{-3}\times10^{23}~\mathrm{cm}^{-2}$.}
   \footnotetext[7]{Luminosity corrected assuming \citet{Panessa06} results. See Sect.~\ref{sec:xdsources} for details.}
 \end{minipage}
\end{table*}

\subsection{Source by source analysis}
\label{sec:xdsources}

\subsubsection*{IRAS 00182-7112}
This type 2 QSO was detected only in the EPIC-PN camera. We modeled the
spectrum of IRAS 00182-7112 with a reflection component, using the
\texttt{pexrav} model \citep{pexrav}. The photon index is fixed to 2
and a pure reflection component is assumed. This implies a lower limit
to the column density of the absorber material
($N_H>10^{25}~\mathrm{cm}^{-2}$).  We marginally detect a narrow
emission line at $6.75^{+0.08}_{-0.11}$~keV (significance
$2-3\sigma$), consistent with He-like Fe line (this energy it is also
consistent at $2\sigma$ level with neutral Fe 6.4~keV line).

\textsl{ISO} and \textsl{Spitzer} IR data suggest the presence 
of a deeply obscured nuclear power source \citep{Tran01,Spoon04}. This
is qualitatively consistent with our X-ray analysis results. The
X-rays detected by XMM-\textsl{Newton} should be the reflected
emission from the AGN, and the line is an iron K$_\alpha$ fluorescent
emission from the reflecting material. The equivalent width of this
line, $0.8\pm0.6$~keV, is consistent with the CT hypothesis, but the
poor quality of the spectrum prevents us from reaching any stronger
conclusion. Assuming that the direct X-ray emission is completely
absorbed by CT material, we have found that the intrinsic 2-10~keV
luminosity of the AGN responsible for the reflection component is
$6.3\times10^{44}(2\pi/\Omega_{refl})~\mathrm{erg~s}^{-1}$, where
$\Omega_{refl}<2\pi$ is the solid angle subtended by the reflector at
the illuminating source. This X-ray emission gives an estimate
to the bolometric luminosity of the AGN of
$\sim1.1\times10^{13}L_{\odot}$, which is consistent with the IR
observations (see Figs.~\ref{fig:povsfir},~\ref{fig:RRvsOBS}).

In conclusion, our X-ray analysis of this source points to an AGN with
CT obscuration.

\subsubsection*{IRAS F00235+1024}
XMM-\textsl{Newton} observed this NL SB galaxy for 26 ks, but it was
not detected by the EPIC cameras. \citet{Wilman03}, assuming a thermal
\texttt{mekal} model ($kT=0.5$~keV), estimate an upper limit to the
0.5-2~keV luminosity of $2.8\times10^{42}~\mathrm{erg~s}^{-1}$,
consistent with our result in this band. The limit in the soft band
implies a weak SB emission in X-rays, compared to that expected from
the IR luminosity (\citealt{Farrah02submm}, see
Table~\ref{tab:lumin}). Their upper limit to the hard X-ray luminosity
($\sim 1.9\times 10^{44}~\mathrm{erg~s}^{-1}$) is also consistent with
ours.

\subsubsection*{IRAS 07380-2342}
This NL object has not been detected by XMM-\textsl{Newton}.

\subsubsection*{IRAS 09104+4109}
The type 2 QSO  IRAS 09104+4109 resides in a rich cluster
\citep{Klein88}. The X-ray soft extended emission from the ICM was
already detected by \textsl{ROSAT} \citep{Fabian95}. We subtracted
this foreground as explained in Section \ref{sec:clsubtraction}.

The source spectra were extracted from a circular region of $20''$ for
all detectors, and the cluster spectra from an annular region between
$20''$ and $90''$ (constrained to the CCD where the source is
located). The cluster emission was fitted with a \texttt{mekal}
model. The temperature is $kT=5.5\pm0.4$~keV, and the metal abundance
is $0.30\pm0.12$ Z$_{\odot}$, which is consistent with the mean Fe
abundance for clusters with a temperature greater than 5 keV
\citep{Horner01}. We obtained the radial brightness profile using
\textsl{Chandra} data: a $\beta$-model with a core radius of $4''.6\pm0''.6$ 
and $\beta = 0''.576^{+0''.018}_{-0''.016}$.

The cluster emission represents 62\% of the total 0.5-10~keV
luminosity. The spectrum of the source was fitted with a power law with
a Fe K$\alpha$ broad ($\sigma = 0.27\pm0.09$~keV) emission line at
$6.61^{+0.08}_{-0.10}$~keV. This broad line could be explained as a
complex of neutral and ionized narrow lines merged due to the low
resolution of the detector.

Although IRAS 09104+4109 is classified as a type 2 QSO, the
XMM-\textsl{Newton} data did not reveal any intrinsic absorption
feature. However, the \textsl{Chandra} observation of this source
suggested a column density of $3\times10^{23}~\mathrm{cm}^{-2}$
\citep{Iwasawa01}. Also, \textsl{Beppo}SAX detected this object at
energies greater than 10 keV, pointing to non-thermal quasar emission
emerging from a thick absorbing torus \citep{Fran00}, with
$N_H\sim7\times10^{24}~\mathrm{cm}^{-2}$ .

Our combined analysis of the \textsl{Beppo}SAX and XMM-\textsl{Newton}
data sets (which will be taken as the best fit for this source in what
follows) shows that a reflection-only model is needed to explain the
complete spectrum in the 0.2 to 50~keV range. This implies a lower
limit ($N_H>10^{25}~\mathrm{cm}^{-2}$) to the column density of the
absorber, which it is consistent with \citet{Iwasawa01}, where a similar
analysis of this source is done with \textsl{Beppo}SAX and
\textsl{Chandra} observations. Assuming $\Gamma=1.4$, which is in 
the flatter side of the photon index distribution of quasars
\citep{Reeves00}, they found that a cold reflection model without
transmitted component fits well the complete data.

This model gives an intrinsic hard X-ray luminosity of 
$2\times10^{45}(2\pi/\Omega_{refl})~\mathrm{erg~s}^{-1}$ for the
AGN. The estimated bolometric luminosity of this source is then
$\sim3.5\times10^{13}L_{\odot}$, which is consistent with its IR data (see
Figs.~\ref{fig:povsfir},~\ref{fig:RRvsOBS}).

We have also found in our combined analysis a new thermal component,
but its temperature ($kT\sim3$~keV) is too high to be associated to SB
or AGN emission. Since there is no evidence of a SB component in the
IR SED of this source \citep{Rowan00}, this thermal component is
probably due to an incomplete subtraction of the cluster
emission. Previous results indicate that a strong cooling flow of the
ICM is taking place in the core of the cluster
\citep{Fabian95,Allen98,Ettori99,Iwasawa01}, so the isothermal ICM
hypothesis we have assumed could underestimate the cluster luminosity
in the central region.

We can confirm that IRAS 09104+4109 is probably a CT source, and that
the X-ray emission detected below 10 keV by XMM-\textsl{Newton} is
only a reflection continuum from cold matter.

\subsubsection*{PG 1206+459}
The XMM-\textsl{Newton} observation of this QSO is contaminated by
background flares at the beginning and at the end of the
observation. The final effective exposure is $\sim7$~ks. The spectra
continuum has been modeled with a power law. We have not detected
significant intrinsic absorption or soft excess. The lack of
absorption in the X-ray spectra is consistent with the optical and IR
evidences \citep{Haas98,Rowan00}. Its hard X-ray luminosity is
$1.3\times10^{45}~\mathrm{erg~s}^{-1}$, which gives a bolometric
luminosity of $\sim2.2\times10^{13}L_{\odot}$, in agreement with the
IR data (see Figs.~\ref{fig:povsfir},~\ref{fig:RRvsOBS}). The X-ray
spectrum of this object is therefore consistent with having a pure AGN
origin.

\subsubsection*{PG 1247+267}
The X-ray spectrum of this QSO has been modeled as a power law and a
thermal component.  A \texttt{pexrav} model is formally the best fit
($\chi^2/\nu=224/282$) of these data. However, the photon index
obtained ($\Gamma\sim2.3$) with this model is slightly larger than the
one expected for an AGN. Moreover, the reflection scaling factor
($\sim4$) is quite larger than the typically expected for type 1
sources (within 0 and 1). No other reflection features have been found
in the X-ray spectrum. Therefore, we have adopted the power law plus
thermal component as our best fit. The bolometric luminosity
of the AGN is $\sim1.4\times10^{14}L_{\odot}$, using its hard X-ray
luminosity. This result is consistent with the IR observations (see
Figs.~\ref{fig:povsfir},~\ref{fig:RRvsOBS}).

The temperature of the thermal component
($kT=0.48^{+0.23}_{-0.17}$~keV) is consistent with the typical
temperature of a SB galaxy. However, the bolometric luminosity that we
can estimate through the soft X-ray emission for this SB is
$\sim10^{49}~\mathrm{erg~s}^{-1}$, much higher than the
\citet{Rowan00} estimate ($<5.2\times10^{46}~\mathrm{erg~s}^{-1}$). 
Therefore, the soft excess component is too luminous to have a pure 
SB origin. Furthermore, its soft excess-to-power law soft X-ray
luminosity ratio is $\sim0.4$, which is typical for soft excess
observed in AGN. The X-ray spectrum of this source is consistent with
being dominated by an AGN.

\subsubsection*{IRAS F12509+3122}
A significant fraction ($\sim50\%$) of the observation time of this
QSO is affected by high background. The PN and MOS spectra can be
fitted by a power law model and a thermal component with
$kT=0.21\pm0.03$~keV, at a lower energy than that expected for a
standard SB, but consistent with soft excess originated in an AGN. The
bolometric luminosity inferred from the X-ray thermal luminosity is
one order of magnitude greater than the estimate for a SB using IR
data \citep{Farrah02submm}. The thermal-to-power law luminosity ratio
is $\sim1.2$, in the range of AGN soft excess. The X-ray spectrum of
this source is also AGN-dominated.

\begin{table*}
 \begin{minipage}[!h]{\textwidth}
 \caption{Flux and luminosity data of the sample.}
 \label{tab:lumin}
 \centering
 \renewcommand{\footnoterule}{}  
  \begin{tabular}{@{}lccccccrrrrr@{}}
  \hline \hline
      &  & \multicolumn{4}{c}{Flux density (Jy)\footnote{Observed by \textsl{IRAS} (from NED).}} & IR & \multicolumn{5}{c}{$\log L$ (erg s$^{-1}$)} \\
  \cline{3-6} \cline{8-12}
       Source   & R\footnote{UK-R magnitude (SuperCOSMOS Sky Survey).}  & 12$\mu$m & 25$\mu$m & 60$\mu$m & 100$\mu$m & Model\footnote{AGN (A) and/or starburst (S) components needed to fit the IR SED (as in Table~\ref{tab:sample}, col.~3). First letter indicates the dominant component.} & $L_{FIR}$$^d$ & $L_{IR}$\footnote{Infrared luminosities in the $40-500~\mu m$ (FIR) and $1-1000~\mu m$ (IR) bands, computed using {\em IRAS} fluxes \citep{Sanders96}.}  
                & $L_{IR,RR}$ & $L_{IR,RR}^{AGN}$ & $L_{IR,RR}^{SB}$ \\
  \hline
 \scriptsize{IRAS 00182-7112 }     &  17.738 & $<$0.06025      &  0.133$\pm$0.010 & 1.20$\pm$0.08 &  1.19$\pm$0.12 & S+A &    46.49 & $<$46.72 & $<$46.93 & $<$46.48 &    46.74 \\
 \scriptsize{IRAS F00235+1024}     & $>$21.5 & $<$0.173        & $<$0.193         & 0.43$\pm$0.06 & $<$0.94        & S+A & $<$46.76 & $<$47.27 &    46.74 &    46.44 &    46.45 \\
 \scriptsize{IRAS 07380-2342 }     &  16.869 &  0.48$\pm$0.03  &   0.80$\pm$0.08  & 1.17$\pm$0.09 &   3.5$\pm$0.3  & A+S &    46.56 &    47.08 &    46.97 &    46.79 &    46.48 \\
 \scriptsize{IRAS 09104+4109 }     &  17.819 &  0.13$\pm$0.03  &  0.334$\pm$0.013 & 0.53$\pm$0.04 & $<$0.44        &  A  & $<$46.42 & $<$46.99 & $<$46.92 &    46.84 & $<$46.15 \\
 \scriptsize{PG 1206+459}~$^d$     &  15.135 &  0.21$\pm$0.04  & $<$0.113	  & 0.26$\pm$0.05 &  0.35$\pm$0.07 &  A  &    47.20 & $<$47.94 & $<$47.80 &    47.78 & $<$46.57 \\
 \scriptsize{PG 1247+267}~$^d$     &  14.621 & $<$0.126        & $<$0.113         & 0.24$\pm$0.05 &  0.17$\pm$0.03 &  A  &    47.70 & $<$48.39 & $<$47.94 &    47.91 & $<$46.76 \\
 \scriptsize{IRAS F12509+3122}     &  16.590 & $<$0.106        &   0.10$\pm$0.03  & 0.22$\pm$0.04 & $<$0.675       & A+S & $<$46.86 & $<$47.37 &    47.00 &    46.76 &    46.62 \\
 \scriptsize{IRAS 12514+1027 }     &  17.654 & $<$0.0632       &  0.190$\pm$0.016 & 0.71$\pm$0.06 &  0.76$\pm$0.15 & S+A &    46.18 & $<$46.51 &    46.63 &    46.27 &    46.39 \\
 \scriptsize{IRAS 13279+3401 }     &  15.689 & $<$0.0937       & $<$0.126         & 1.18$\pm$0.08 &  1.20$\pm$0.18 & A+S &    46.58 & $<$46.84 &    46.53 &    46.36 &    46.04 \\
 \scriptsize{IRAS 14026+4341 }     &  15.651 &  0.12$\pm$0.03  &  0.285$\pm$0.014 & 0.62$\pm$0.06 &  0.99$\pm$0.24 & A+S &    46.26 &    46.70 &    46.54 &    46.34 &    46.11 \\
 \scriptsize{IRAS F14218+3845}     & $>$21.5 & $<$0.0969       & $<$0.075	  & 0.57$\pm$0.06 &  2.10$\pm$0.17 & S+A &    47.80 & $<$48.06 &    46.86 &    46.15 &    46.76 \\
 \scriptsize{IRAS F15307+3252}     &  19.131 & $<$0.065        &  0.071$\pm$0.019 & 0.23$\pm$0.04 & $<$0.71        & A+S & $<$47.07 & $<$47.46 &    47.22 &    47.05 &    46.73 \\
 \scriptsize{IRAS 16347+7037}~$^e$ &  13.979 & 0.059$\pm$0.010 &  0.122$\pm$0.004 & 0.27$\pm$0.05 &  0.35$\pm$0.07 & A+S &    47.42 &    47.86 &    47.81 &    47.73 &    47.04 \\
 \scriptsize{IRAS 18216+6418 }     &  13.403 & $<$0.238        &  0.445$\pm$0.012 & 1.24$\pm$0.05 &  2.13$\pm$0.17 & A+S &    46.49 & $<$46.89 &    46.78 &    46.54 &    46.37 \\
  \hline
  \end{tabular}
 \footnotetext[5]{60 and 100 $\mu$m fluxes are \textsl{ISO} data from \citet{Haas00}.}
 \end{minipage}
\end{table*}

\subsubsection*{IRAS 12514+1027}
This Seyfert 2 galaxy was also observed by \textsl{ROSAT} \citep{Wilman98},
but only XMM-\textsl{Newton} has been able to detect its X-ray
emission. \citet{Wilman03} considered only the PN spectrum, modeled
with a reflection component (\texttt{pexrav}, with fixed $\Gamma=2$)
and a thermal component (\texttt{mekal},
$kT=0.31^{+0.13}_{-0.05}$~keV) corrected by intrinsic absorption
($N_H=1.3^{+0.9}_{-0.7}\times10^{21}~\mathrm{cm}^{-2}$). The resulting
$\chi^2/\nu$ is 8.5/12.

In our analysis, the three EPIC spectra were coadded (see
Sect.~\ref{sec:reduction}). We modeled the spectrum with an absorbed
power law with a photon index fixed to 2, and a thermal component.

The temperature of the thermal component is
$kT=0.35^{+0.17}_{-0.07}$~keV, consistent with that usually observed
in a SB galaxy. The soft component-to-power law luminosity ratio is
slightly lower than that associated to an AGN. The 0.5-2.0~keV
luminosity of this soft component is
$1.5\times10^{42}~\mathrm{erg~s}^{-1}$, which implies a bolometric
luminosity for the SB of
$\sim1.5\times10^{46}~\mathrm{erg~s}^{-1}$. The IR luminosity of the
SB estimated from the analysis of the IR SED of the source is
$\sim5\times10^{46}~\mathrm{erg~s}^{-1}$ \citep{Rowan00}. This thermal
component could be consistent with a SB origin.

The optical, IR and X-ray data point to a CT quasar and a SB 
of comparable bolometric luminosities.

\subsubsection*{IRAS 13279+3401}
This QSO has not been detected by XMM-\textsl{Newton}.

\subsubsection*{IRAS 14026+4341}
The XMM-\textsl{Newton} observation of this QSO 1.5 is heavily
contaminated by background flares. All the PN data are affected by
count rate background greater than 15 counts per second. The MOS data
have a brief interval free of flares, but the source is not
detected.

\subsubsection*{IRAS F14218+3845}
The QSO IRAS F14218+3845 was observed by XMM-\textsl{Newton} in two
occasions. The second observation was heavily affected by high
radiation background. We co-added the six spectra of the different
observations to increase the S/N ratio. The spectrum was modeled with
a power law. No significant soft excess or intrinsic absorption was
found.

The IR data suggest that this HLIRG is a SB dominated source
\citep{Verma02,Farrah02submm}, but our analysis of its
XMM-\textsl{Newton} X-ray spectrum does not reveal any SB
features. Using the upper limit that we have estimated (see
Table~\ref{tab:xraydata}) for a thermal component, the total SB
luminosity is less than $6\times10^{47}~\mathrm{erg~s}^{-1}$, which is
consistent with the SB luminosity estimated by \citet{Farrah02submm}
through IR and sub-mm data ($6\times10^{46}~\mathrm{erg~s}^{-1}$). 
Although a SB component cannot be excluded, the data point to an
AGN-dominated X-ray emission \citep{Fran03}.

\subsubsection*{IRAS F15307+3252}
Previous observations with \textsl{ROSAT} and \textsl{ASCA} detected
no X-ray emission from this QSO 2 \citep{Fabian96,Ogasaka97}. We
detected a faint X-ray emission in the XMM-\textsl{Newton} public
data. The observation of this source is affected by high background
flares. The three EPIC extracted spectra were coadded to increase S/N
ratio. We fitted this spectrum using a power law. We were not able to
find any absorption feature or thermal emission.

XMM-\textsl{Newton} has observed IRAS F15307+3252 on two more
occasions, but the data are still private. \citet{Iwasawa05}, using
the complete data set, found a prominent Fe K$\alpha$ line at
$\sim6.5$~keV, indicating the presence of a CT AGN. This is in
agreement with optical spectropolarimetry data indicating the presence
of a dust-enshrouded quasar \citep{Hines95}. The estimate of the AGN
bolometric luminosity using the observed emission line luminosity is
also consistent with previous results
\citep{Yun98,Aussel98,Verma02,Peeters04}. The hard X-ray emission
detected by us is probably reflected radiation, because of CT
obscuration. \citet{Panessa06} found that the ratio between the
intrinsic and the observed X-ray luminosity in CT Seyfert galaxies is
$\sim60$. We have corrected the X-ray luminosity calculated with the
public XMM-\textsl{Newton} data by this factor.  The resulting hard
X-ray luminosity ($\sim3.2\times10^{45}~\mathrm{erg~s}^{-1}$) is
consistent with the estimate given by \citet{Iwasawa05}
(L$_X>1\times10^{45}~\mathrm{erg~s}^{-1}$), using the luminosity of
the iron emission line. Assuming an AGN origin, this X-ray emission
gives a bolometric luminosity for this source of $\sim5.5\times10^{13}L_{\odot}$, 
in agreement with the IR data (see Figs.~\ref{fig:povsfir},~\ref{fig:RRvsOBS}).

\citet{Iwasawa05} also found extended soft emission, with 
$kT=2.1^{+0.6}_{-0.4}$~keV. They identify this extended emission with
hot gas associated with a relatively poor cluster around this
object. Although no galaxy cluster has been found associated to this
source, an \textsl{HST} observation shows a moderate galaxy
over-density. Moreover, its bolometric luminosity to temperature
relation would be similar to that typical of poor clusters
\citep{Fukazawa04}.

In summary, the X-ray spectrum of this source is consistent with the
emission originated in a CT AGN.

\subsubsection*{IRAS 16347+7037}
The spectrum of the QSO IRAS 16347+7037 was modeled with a power law
and a thermal component. No intrinsic absorption is detected.

The XMM-\textsl{Newton} X-ray spectrum is consistent with a type 1
AGN spectrum, as the optical \citep{Evans98} and IR
\citep{Haas98,Farrah02hst} observations suggest. Previous X-ray data
from \textsl{ASCA} was also consistent, and there was no evidence of iron
K$\alpha$ emission feature or any absorption edge \citep{Nandra95}.

The soft excess has
$L(0.5-2.0~\mathrm{keV})=5.6\times10^{45}~\mathrm{erg~s}^{-1}$. This
would imply a SB bolometric luminosity three orders of magnitude
greater than the SB luminosity calculated with the IR data
(\citealt{Farrah02hst}, see Table~\ref{tab:lumin}). Therefore this
component is unlikely to be associated to a SB. Moreover, its
thermal-to-power law luminosity ratio is consistent with a soft excess
from the AGN. 

This model gives a bolometric luminosity for the AGN of
$\sim2.2\times10^{14}L_{\odot}$, in agreement with the IR
observations of this source (see Figs.~\ref{fig:povsfir},~\ref{fig:RRvsOBS}).
Our analysis points to a pure AGN origin for the X-ray spectrum, in
agreement with the optical and IR data.

\subsubsection*{IRAS 18216+6418}
The PN data of this QSO 1.2 were heavily affected by pile-up, so we
used only the MOS data. The source spectra were extracted from a
20$''$ radius circular region in both MOS detectors. This QSO is
located in a rich cluster, and \textsl{ROSAT} detected the ICM thermal
emission \citep{Saxton97,Hall97}. We subtracted the soft X-ray
emission from the cluster, as explained in
Sect.~\ref{sec:clsubtraction}.

The MOS cameras operated in small-window mode, so we considered the PN
image to model the cluster (the pile-up only affects the central
region of the source). We extracted a spectrum from an annular region
between 20$''$ and 80$''$ (constrained to the CCD where the source is
located). The resulting temperature of the \texttt{zbremss} model was
$kT=2.3^{+1.0}_{-0.6}$~keV. We used also the radial X-ray brightness
profile published by \citet{Fang02}, (core radius of
$17''.6\pm0''.17$, $\beta = 0''.74^{+0''.05}_{-0''.03}$), to
renormalize the cluster model. The cluster X-ray emission is 32\% of
the total 0.5-10~keV luminosity.

The source spectrum best fit is a power law
($\Gamma=1.57^{+0.10}_{-0.11}$) with a soft thermal
component. A \texttt{pexrav} model is formally the best fit
($\chi^2/\nu=321/333$) of this spectrum. As discussed in PG 1247+267,
the steeper photon index ($\Gamma\sim2.3$) and the reflection scaling
factor $\gg1$ ($R\sim15$) lead us to adopt the power law plus thermal
component as our best fit.

The photon index of the power law is not consistent with previous
X-ray observations with \textsl{ASCA} ($\Gamma = 1.75\pm0.03$,
\citealt{Yamas97}) and \textsl{Chandra} ($\Gamma =
1.761^{+0.047}_{-0.052}$, \citealt{Fang02}).

In Fig.\ref{fig:iras18216} a systematic effect in the $\Delta\chi^2$
spectrum can be seen above 5~keV. This may be due to a bad extraction
of the cluster emission or to the use of the blank-field
background. If we ignore the data above 4.5~keV, we get a steeper
photon index ($\Gamma = 1.68\pm0.11$), compatible with previous X-ray
observations.

A thermal component is also detected, with a temperature of
$0.49^{+0.09}_{-0.08}$~keV, consistent with SB emission. However, if
this emission were associated to the SB, the bolometric luminosity of
the SB would be three orders of magnitude higher than the luminosity
calculated using the IR data \citep{Farrah02submm}. The soft
component-to-power law luminosity ratio is in the range of that
typically observed in an AGN.

\textsl{Ginga} \citep{Kii91}, \textsl{ASCA} \citep{Yamas97} and 
\textsl{Chandra} \citep{Fang02,Yaqoob05} detected iron emission features 
in this HLIRG. \citet{Jimenez07} detected a Fe-K emission line
with a complex structure in the PN XMM-Newton spectrum of this
source. We have also detected an emission line in the 6-7 keV rest
frame band, but the significance of the detection ($<2\sigma$) was
below our adopted threshold, and therefore we have not considered it
further. We have estimated a $3\sigma$ flux upper limit to a broad
($\sigma=0.1$~keV) line component at 6.4~keV of
$<3\times10^{-5}~\mathrm{photons~cm}^{-2}~\mathrm{s}^{-1}$, consistent
with the value
$\sim(3\pm1)\times10^{-5}~\mathrm{photons~cm}^{-2}~\mathrm{s}^{-1}$
obtained with the \textsl{Chandra} data \citep{Fang02}.

Assuming an AGN origin, the hard X-ray emission gives a 
bolometric luminosity for this source of
$\sim7\times10^{13}L_{\odot}$, in agreement with its IR data (see
Figs.~\ref{fig:povsfir},~\ref{fig:RRvsOBS}). The X-ray spectrum of
this object is consistent with a pure AGN origin.

\section{Discussion}
\label{sec:discuss}
In their study of X-ray emission from ULIRGs, \citet{Fran03} define
that the X-ray emission of a ULIRG is AGN-dominated if it presents
either: a) a high X-ray luminosity,
$L(2-10~\mathrm{keV})>10^{42}~\mathrm{erg~s}^{-1}$; b) a heavily
obscured hard X-ray component with $N_H > 10^{22}~\mathrm{cm}^{-2}$
(very flat or inverted hard X-ray spectra); or c) a Fe-K emission
complex at $\sim6.4$~keV with equivalent width $\gtrsim1$~keV (iron
fluorescent emission from material illuminated by the AGN). The ten
detected HLIRGs from our sample present at least one of the three
characteristics above, thus showing an AGN-dominated X-ray spectrum.

This result is in agreement with the trend noted by \citet{Veilleux99}
for ULIRGs: the fraction of sources with Seyfert characteristics
increase with L$_{IR}$ (from $\sim25$\% among ULIRGs with
L$_{IR}<10^{12.3}\mathrm{L}_{\odot}$ to $\sim50$\% among those with
L$_{IR}>10^{12.3}\mathrm{L}_{\odot}$). However, we must keep in mind
that our sample is not complete, and could be slightly biased towards
AGN. The first subsample in \citet{Rowan00}, which is not biased in
favour of AGN (see Sect.~\ref{sec:sample}), presents 50\% of objects
with Seyfert characteristics.

Five objects from our sample (four type 2 objects and one NL object)
are probably CT, as reported in the literature. Our analysis of the
XMM-\textsl{Newton} data of two sources (IRAS 00182-7112, IRAS
12514+1027) are consistent with the CT hypothesis, as well as our
combined analysis of XMM-\textsl{Newton} and \textsl{Beppo}SAX data
for IRAS 09104+4109. IRAS F00235+1024 has not been detected, probably
because it is heavily obscured.  We have found no absorption features
in IRAS F15307+3252, but as explained in Sect.~\ref{sec:xdsources},
recently published results from XMM-\textsl{Newton} private data are
consistent with CT absorption \citep{Iwasawa05}.

\begin{figure}[t]
\centering
\includegraphics[angle=-90, width=\linewidth]{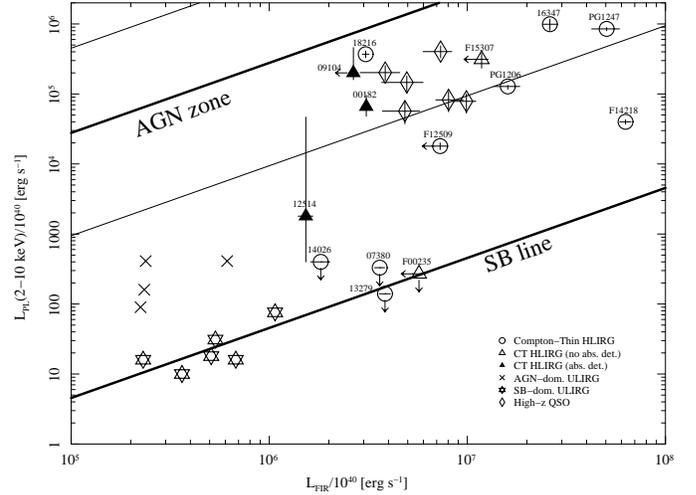}
\caption{2-10~keV X-ray luminosity of the power law component versus 
FIR luminosity (40-500~$\mu$m). Filled symbols represent sources
where we have detected X-ray absorption (models D or F in
Table~\ref{tab:xraydata}). The top thick solid line and area between
the thin solid lines (``AGN-zone'') indicates the X-ray luminosity
expected for an AGN of a given FIR luminosity \citep[90\%
dispersion,][]{Elvis94,Risaliti04}. The bottom thick solid line
(``SB-line'') indicates the X-ray luminosity expected for a SB of a
given FIR luminosity \citep{Persic04,Kenn98}. See
Sect.~\ref{sec:discuss} for details.}
\label{fig:povsfir}
\end{figure}

We have calculated the FIR (40-500~$\mu$m) luminosities ($L_{FIR}$)
using the \textsl{IRAS} fluxes (\citealt{Sanders96}, see Table
\ref{tab:lumin}, column 7). In Fig.~\ref{fig:povsfir} we have plotted
the 2-10~keV luminosity of the power law component for each source
versus the FIR luminosity. We have included the ULIRGs data from
\citet{Fran03} and the high-$z$ QSO from \citet{Stevens05} for
comparison. No correlation between the 2-10~keV and the FIR luminosity is
found in HLIRGs, although it must be kept in mind that the sample is
not complete in any sense.

We have estimated the expected X-ray luminosity for a standard AGN,
given its FIR luminosity. To this end, we calculated the
2-10~keV-to-FIR luminosity ratio typical of nearby bright QSOs, using
their mean SED. Since the original \citet{Elvis94} sample was biased
towards sources with high X-ray luminosity, we have employed the
\citet{Risaliti04} new data on QSO SED. Using this corrected SED, the
ratio of the 2-10~keV band to bolometric luminosity changes
significantly by a factor of 2, from $\sim0.03$ to $\sim0.015$.  The
X-ray luminosity derived from the IR luminosity using the latter ratio
is plotted in Fig.~\ref{fig:povsfir} with a thick solid line. The top
area between thin lines (``AGN-zone'') is the dispersion of the SED,
calculated with the 90 percentile distribution \citep{Elvis94}.

We can also calculate a relationship between FIR and X-ray luminosity
for SB galaxies. The SFR of a SB is given by its FIR luminosity by
SFR$_{FIR}\sim L_{FIR}/2.2\times10^{43}$~M$_\odot$~yr$^{-1}$
\citep{Kenn98}, and by its 2-10~keV X-ray luminosity by SFR$_X\sim
L_{2-10}/10^{39}$~M$_\odot$~yr$^{-1}$ (\citealt{Persic04}, all
luminosities are in CGS units). Assuming equal SFR, the 2-10~keV to
FIR luminosity ratio is $L$(2-10~keV)$/L_{FIR}\sim4.5\times10^{-5}$. 
This relation is shown in Fig.~\ref{fig:povsfir} as the lower thick
solid line (``SB-line'').

\begin{figure}[t]
\centering
\includegraphics[angle=-90, width=\linewidth]{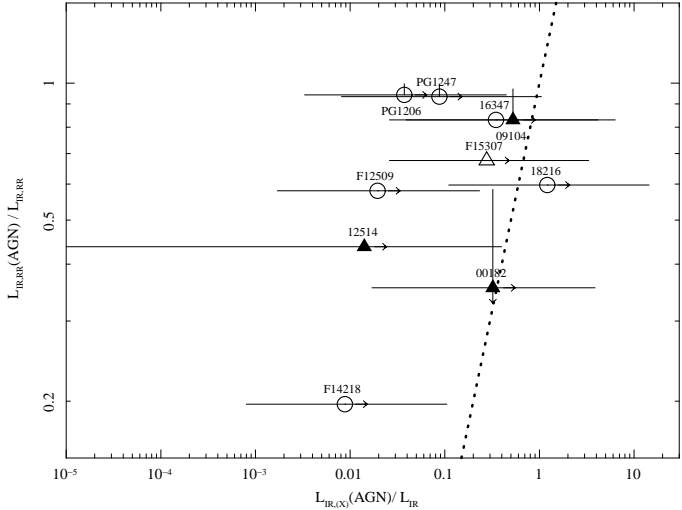}
\caption{AGN to total-IR luminosity ratio, calculated by \citet{Rowan00},
versus AGN (estimated through our X-ray data) to total (computed with
the \textsl{IRAS} fluxes) IR luminosity ratio. The dotted line means
equal values for the ratios. In case of upper limits we plot as error
bars the dispersion of X-ray-to-IR luminosity ratio from
\citet{Elvis94}. IRAS~F00235+1024, IRAS~07380-2342, IRAS~F12509+3122,
IRAS~12514+1027, IRAS~13279+3401, IRAS~14026+4341 and
IRAS~F14218+33845 IR data are taken from the \citet{Farrah02submm} results;
IRAS~F15307+3252 IR data is taken from the \citet{Verma02} results;
IRAS~16347+7037 IR data is taken from the \citet{Farrah02hst} results. 
Symbols as in Fig.~\ref{fig:povsfir}.}
\label{fig:RRvsOBS}
\end{figure}

Most HLIRGs and all high-$z$ QSO are in the ``AGN-zone'', while only
the AGN-dominated ULIRGs and two HLIRGs seem to be composite sources:
their X-ray luminosity is too high to be produced only by a SB (above
the ``SB-line''), and their FIR luminosity is too high to be produced
only by an AGN (to the right of the ``AGN-zone''). The SB-dominated
ULIRGs are concentrated near the ``SB-line''. The upper limits for the
X-ray-undetected HLIRGs seem to indicate that their hard X-ray emission
comes only from SB activity.  However, the X-ray emission of the
non-detected sources could be affected by heavy obscuration. Actually,
one of the non-detected sources, IRAS F00235+1024, is probably CT as
seen from their IR spectrum so its X-ray emission could be depressed
by heavy absorption. However the remaining sources are optical QSOs,
where little or no absorption is expected. For example, the QSOs from
\citet{Stevens05} show relatively low absorption ($21<\log
N_H<22~\mathrm{cm}^{-2}$). Recent XMM-\textsl{Newton} observations of these
high-$z$ HLIRGs suggest highly ionized winds with $22.5<\log
N_H<23.5~\mathrm{cm}^{-2}$ \citep{Page07}. Further sensitive data on
isotropic indicators (such as FIR or MIR or $>10$~keV emission) are
needed to investigate the seemingly contradictory nature of these
HLIRGs.

Note that if the \citet{Elvis94} X-ray-to-FIR ratio is used instead of the
\citet{Risaliti04} ratio, only three HLIRGs (IRAS 09104+4109, IRAS
16347+7037 and IRAS 18216+6418) would lie on the ``AGN-zone'', and the
rest would be considered as composite AGN/SB sources. This confirms
the relevance of the \citet{Risaliti04} correction.

The HLIRGs FIR emission is systematically above the \citet{Risaliti04}
estimate for a standard local QSO of the same X-ray luminosity (i.e.,
the sources are located on the right of the thick upper line in
Fig.~\ref{fig:povsfir}). This FIR excess could be associated to the
SB activity in HLIRGs. Alternatively, this could also hint to a
possible difference between the standard QSO SED and the AGN component
of the HLIRGs SED. In this line, it has been shown that the shape of
the SED is probably related to the luminosity \citep{Marconi04}.

\begin{figure}[t]
\centering
\includegraphics[angle=-90, width=\linewidth]{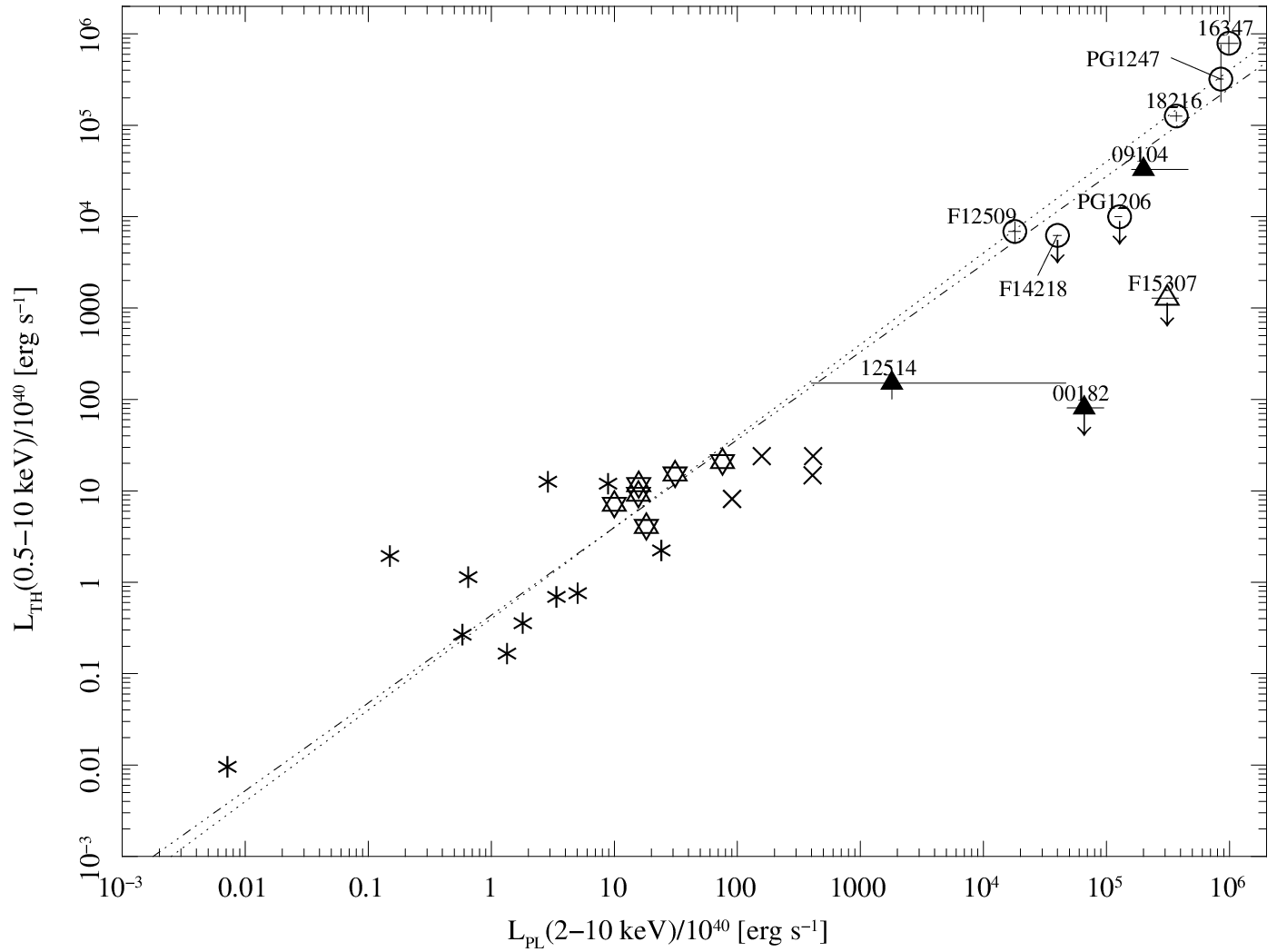}
\caption{L$_X$ (0.5-10~keV) of the soft excess component versus
L$_X$ (2-10~keV) of the power law component. Asterisks represent
SB galaxies from \citet{Fran03}. The dotted line is a correlation
obtained by \citet{Fran03} for SB-dominated ULIRGs. The dash-dotted
line is a correlation for the SB-dominated ULIRGs and the SB galaxies
samples calculated by us. The soft excess in IRAS 09104+4109 is
probably associated to cluster emission (see Sect.~\ref{sec:xdsources}
for details). Symbols as in Fig.\ref{fig:povsfir}.} 
\label{fig:thvspo}
\end{figure}

We tried to unravel the origin of the excess infrared emission with
respect to that predicted using the \citet{Risaliti04} QSO SED. We
have estimated the AGN contribution to the total IR luminosity of the
HLIRGs in two independent ways, in order to know if this IR excess
comes from SB activity, or from an intrinsic difference in the AGN
SED.

On one hand, \citet{Rowan00} has modeled the IR SED of all sources in
this sample with radiative transfer models, and he estimated the
contribution of the AGN dust torus ($L_{IR,RR}^{AGN}$) and the SB to
the total IR luminosity (1-1000~$\mu$m, $L_{IR,RR}$)
(Table~\ref{tab:lumin}, columns 9 and 10 respectively). We have
corrected the relative contribution of IRAS~F00235+1024,
IRAS~07380-2342, IRAS~F12509+3122, IRAS~12514+1027, IRAS~13279+3401,
IRAS~14026+4341, IRAS~F14218+3845, IRAS~F15307+3252 and
IRAS~16347+7037 with the results of \citet{Verma02},
\citet{Farrah02submm} and \citet{Farrah02hst}, where new IR and sub-mm
data are used.

On the other hand, we have calculated this ratio using the power law
X-ray luminosity. We assume that this emission is associated to an AGN
with a standard SED \citep{Risaliti04}, and we calculate the
IR-to-2-10~keV luminosity ratio, similarly to that done in
Fig.~\ref{fig:povsfir}. We obtain an independent estimate of the
expected IR luminosity of the AGN ($L_{IR,(X)}^{AGN}$).

In Fig.~\ref{fig:RRvsOBS} we compare the relative contribution of the
AGN to the total IR luminosity calculated by \citet{Rowan00}, to our
estimate obtained from X-ray data (we have included only the X-ray
detected sources). The dotted line corresponds to the 1:1 relation,
i.e. an agreement between the two estimates. Most of the
sources have lower limits on the abscissae because their 12 and 25
$\mu$m \textsl{IRAS} fluxes are upper limits (see Table
\ref{tab:lumin}).

Our estimates of the AGN relative contribution for all sources are
formally consistent with that of \citet{Rowan00}. However, there seems
to be a systematically overestimate of the IR AGN component from
\citet{Rowan00} with respect to the X-ray measurements.

The values of $L_{IR,(X)}^{AGN}/L_{IR}$ plotted in
Fig.~\ref{fig:RRvsOBS} are independent of the SB luminosity, so this
disagreement is probably due to the IR-to-X-ray ratio used to estimate
the IR luminosity from the X-ray luminosity. This favours the
hypothesis of an intrinsic difference between the standard QSO SED and
the AGN component of the HLIRGs SED. A detailed radio-to-X-ray study of
the spectral energy distribution of HLIRGs is needed to solve this
question.

\begin{figure}[t]
\centering
\includegraphics[angle=-90, width=\linewidth]{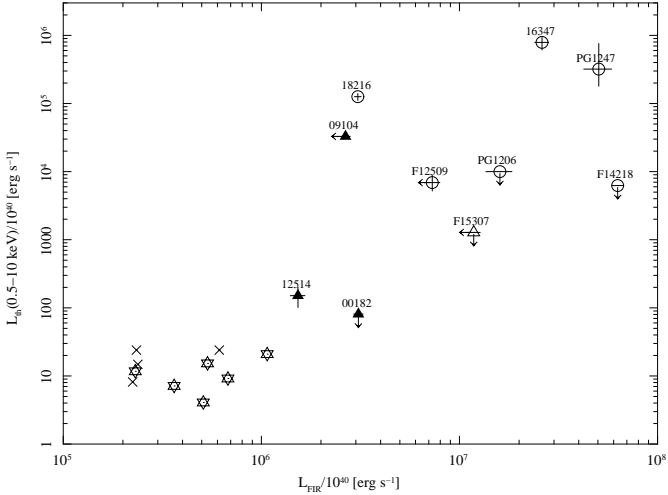}
\caption{L$_X$ (0.5-10~keV) of the soft excess component versus FIR luminosity. 
The soft excess in IRAS 09104+4109 is probably associated to cluster emission (see Sect.~\ref{sec:xdsources}
for details). Symbols as in Fig.\ref{fig:povsfir}.} 
\label{fig:thvsfir}
\end{figure}

Fig.~\ref{fig:thvspo} is a luminosity-luminosity plot of the soft
X-ray excess versus power law components. In IRAS~09104+4109,
IRAS~F12509+3122, PG~1247+267, IRAS~16347+7037 and IRAS~18216+6418 the
soft excess component is too luminous to come only from a SB (see
Sect.~\ref{sec:xdsources} for a detailed description of each
source). In the case of IRAS 09104+4109 the soft excess is probably
due to an incomplete subtraction of the cluster emission, while in the
remaining four sources is probably of the same origin as in luminous QSO
\citep{Piconcelli05}.

An X-ray thermal component associated to SB emission has been observed in all 
ULIRGs from \citet{Fran03}. However, only in 1 out of 14 HLIRGs we have found 
a soft X-ray emission whose origin could be associated to SB activity. 

Oddly, the above HLIRGs with AGN-like soft excess emission follow the
correlation found for SB-dominated ULIRGs by \citet{Fran03} (dotted
line in Fig.~\ref{fig:thvspo}). To increase the statistics and to test
if this correlation holds at lower IR luminosities, we have included a
sample of SB galaxies (\citealt{Fran03} and references therein). We
have calculated a non-parametric correlation coefficient (generalized
Kendall's Tau\footnote{We employed the ASURV software for this test
\citep{asurv1,asurv2}.}) for the SB galaxies and the SB-dominated
ULIRGs, finding $Z_{\tau}=3.69$ with a significance
level\footnote{Note that even excluding the isolated source in the
bottom left corner, this correlation remains almost unchanged.} of
99.98\% ($>3\sigma$). The correlation slope is consistent with that
obtained by \citet{Fran03} for SB-dominated ULIRGs only.

We have investigated the relationship between the X-ray soft excess
component luminosity and the FIR luminosity (Fig.~\ref{fig:thvsfir}),
finding no clear correlation. A test using generalized Kendall's Tau
confirms this impression: $Z_{\tau}=2.13$, with a significance level
96.66\% ($<3\sigma$).

\begin{figure}[t]
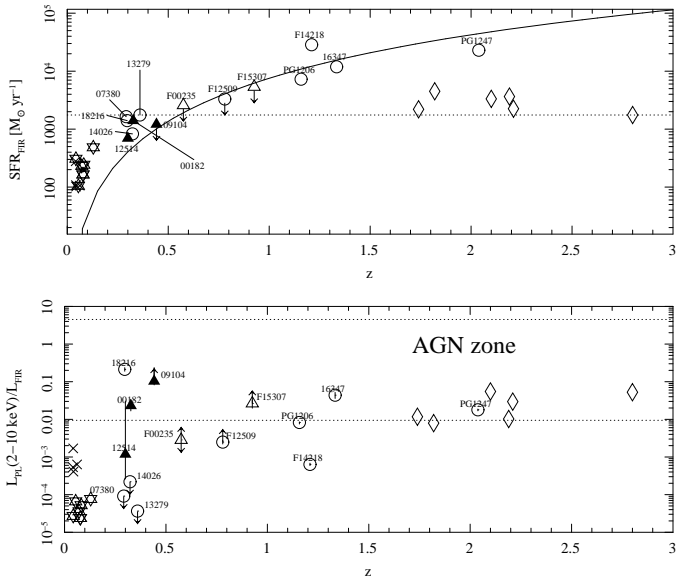

\begin{center}
  \mbox{ \subfigure{
   \includegraphics[angle=-90,width=1\linewidth]{aruiz_XMMhlirgs_fig6a}
   \label{fig:zSFR} } } 
  \mbox{ \subfigure{
   \includegraphics[angle=-90,width=1\linewidth]{aruiz_XMMhlirgs_fig6b}
   \label{fig:zRatio} } }
\end{center}
\caption{Symbols as in Fig.\ref{fig:povsfir}. 
\textbf{Top:} SFR derived from FIR luminosity \citep{Kenn98}, versus redshift. The 
dotted line represents the SFR limit corresponding to the HLIRGs definition. The solid 
line is the SFR for a source with \textsl{IRAS} fluxes equals to the \textsl{IRAS} FSC sensitivity limits 
\citep{irasFSC}. \textbf{Bottom:} 2-10~keV power law-X-ray-to-FIR luminosities ratio versus 
redshift. The area between the dotted lines represent the expected ratio for a quasar (with 
90\% of dispersion)  with a standard SED \citep{Elvis94,Risaliti04}.}
\label{fig:sfrvsz}
\end{figure}

We have also tested a possible cosmological evolution in the
sample. We have estimated the SFR for each source using its IR
luminosity \citep{Kenn98}. Their SFR and 2-10~keV-to-IR-luminosity ratio
versus redshift are plotted in Fig.~\ref{fig:sfrvsz}.  Cosmic star
formation shows an important decline between $z\sim2$ and the present
day \citep{Fran99}, so we expect an increment of the SFR of HLIRGs up
to $z\sim2$. Higher SFR at higher redshift is observed in the upper
panel of the Fig.~\ref{fig:sfrvsz}. However the sources follow the
lower envelope, which is the \textsl{IRAS} FSC sensitivity limit
(solid line in Fig.~\ref{fig:sfrvsz}), indicating clearly a selection
effect. Therefore, we can not draw conclusions about the dependence of
SFR with redshift.

As shown in the bottom panel of the Fig.~\ref{fig:sfrvsz}, the ratio
of hard X-ray-to-FIR luminosity remains constant with $z$. This holds
even if we subtract the FIR luminosity emitted by the AGN, calculated
using the X-ray PL luminosity as in Fig.~\ref{fig:RRvsOBS}.

We have seen that the IR emission is consistent with an AGN origin,
but if we assume that the IR excess shown in Fig.~\ref{fig:povsfir} is
associated to the SB activity, Fig.~\ref{fig:sfrvsz} shows that its
evolution must be similar to that of the X-ray emission. This, in
turn, suggests that the presence of a SB and the occurrence of AGN
activity through accretion onto a super-massive black hole are
physically related. This result is in agreement with the coeval black
hole/stellar bulge formation hypothesis \citep{Granato04,Stevens05,DiMatteo05}.

\section{Conclusions}
\label{sec:conc}
We have performed a systematic X-ray study of a sample of 14
Hyper-Luminous Infrared Galaxies using XMM-\textsl{Newton} data from
the archive, and our own private data. We modeled the X-ray spectra of
each source, finding very heterogeneous spectral properties. Our
results are summarized as follows:
 
\begin{enumerate}
\item All X-ray detected HLIRGs of the sample (ten sources) have
AGN-dominated X-ray spectra.

\item No correlation is found between the 2-10~keV and IR luminosities in
HLIRGs. The hard X-ray luminosity of most (eight) HLIRGs is consistent
with emission from an AGN component only. However two HLIRGs (as well
as all the AGN-dominated ULIRGs) seem to show a composite AGN/SB
nature: their X-ray luminosity is too high to be produced only by a
SB, and their IR luminosity is too high to be produced by only an
AGN. The remaining four HLIRGs are undetected in X-rays.

\item The hard X-ray luminosity associated to the AGN is systematically below the
one expected for a local QSO \citep{Elvis94,Risaliti04} of the same
IR luminosity.  This seems to suggest that there is an intrinsic
difference between the AGN component of the HLIRGs SED and the SED of
local QSOs.  A detailed radio-to-X-ray study of the HLIRGs SED is
needed to understand this issue.

\item There are five Compton-thick candidate sources in our sample,
including all the type 2 AGN (four) and one starburst object
undetected in X-rays. We have found evidence for X-ray absorption in three of
the type 2 AGN. Our analysis supports the Compton-thick nature of
these sources.

\item We have detected five HLIRGs with soft excess emission. Four sources
present luminosities consistent with type 1 AGN soft excess. While all
ULIRGs from \citet{Fran03} present a soft X-ray thermal component
associated to a SB, only in one HLIRGs we have found a soft excess
probably associated to SB emission.

\item The hard-X-ray-to-FIR luminosity ratio remains constant with $z$, 
suggesting that the AGN and SB phenomena are physically connected in HLIRGs.
\end{enumerate}

\begin{acknowledgements}
A.R. acknowledges support from a Universidad de Cantabria fellowship.
F.P. acknowledges support by a ''Juan de la Cierva'' fellowship.
Financial support for A.R., F.P. and F.J.C. was provided by the Spanish
Ministry of Education and Science, under project ESP2003-00812.

This paper is based on observations obtained with XMM-\textsl{Newton},
an ESA science mission with instruments and contributions directly
funded by ESA Member States and NASA; \textsl{Chandra}, a NASA
mission; \textsl{Beppo}SAX, a project of the ASI (Italy) with
participation of the NIVR (Netherlands) and the Space Science
Department of ESA; \textsl{ISO}, an ESA project with the participation
of ISAS and NASA; and \textsl{IRAS}, a joint project of the US, UK and
the Netherlands.  This research also has made use of the SuperCOSMOS
Sky Survey (SSS) data base and the NASA/IPAC Extragalactic Database
(NED) which is operated by the Jet Propulsion Laboratory, California
Institute of Technology, under contract with the National Aeronautics
and Space Administration. 
\end{acknowledgements}


\end{document}